\documentstyle[12pt,epsf]{article}

\begin{document}
 
\begin{center}

{\LARGE
Hamiltonian renormalization for bound state problem
       in gluodynamics\\}
\vskip 20pt
%
{\Large
Elena Gubankova, 
         Chueng-Ryong Ji, and Stephen R. Cotanch\\}
\bigskip
{\it
Department of Physics, North Carolina State University, Raleigh,
     NC 27696-8202 } 

\bigskip
\end{center}

\date{\today}
%
\begin{abstract}

The systematic approach to study bound states in gluodynamics
is presented. The method utilizes flow equations together with
low-energy phenomenology, that provides the perturbative
renormalization scaling in conjuction with the change of the basis 
to constituent gluon states. The renormalized effective Hamiltonian 
of gluodynamics up to the second order is obtained at low energies, 
which provides a kind of constituent gluon model for glueball bound states.
The approach allows to include perturbative QCD corrections
into nonperturbative calculations of many-body techniques.  
The performed numerical calculations support the constituent picture 
of hadronic observables.

\end{abstract}

\section{Introduction}
\label{sec:intr}

QCD is a widely accepted theory of strong interactions, where the phenomenon
of asymptotic freedom provides the success of perturbative calculations.
There is still a gap between perturbative behavior of QCD and its 
low-energy limit, where physical observables are described based on phenomenology.
An ultimate goal of this study is to connect both regimes and obtain
a low-energy effective Hamiltonian from canonical QCD Hamiltonian,
which then has to be solved for bound states.

We suggest a framework which incorporates a perturbative behavior of QCD
and our knowledge from QCD motivated phenomenological models. We consider pure 
gluodynamics in order to disentangle two problems of low-energy QCD;
confinement and chiral symmetry breaking.

Our primary aim is to obtain a renormalized gap equation in the gluon sector, 
which yields a gluon mass gap. The previous study concerns the fermion gap equation
for dynamical quark mass \cite{orsay,adler}. In this case a nonzero chiral condensate 
$\langle q \bar{q}\rangle$ is produced, which leads to a spontaneous 
chiral symmetry breaking.
In gluodynamics the massive gluon mode provides a double pole instead
of a simple one in gluon propagator \cite{chernodub}. This is used as a phenomenological 
criteria of confinement \cite{chernodub,baker,bergerhoff}. 
Actually, we proceed in another way.
We implement a confining potential, which provides in our calculations
nonzero gluon mass and gluon condensate $\langle F_{\mu\nu}F_{\mu\nu}\rangle$.
One encounters a problem of UV-divergences with the Coulomb potential.
For potentials which do not lead to UV-divergences, such as a pure 
confining potential $1/q^4$, unrenormalized gluon gap equation was
obtained by Szczepaniak et.al. \cite{ssjc96}. First, we derive the renormalized
gap equation, which provides a cut-off independent gluon dispersion
relation up to leading order; it is used then in the Tamm-Dancoff bound state equation 
to calculate glueball mass nonperturbatively.

We utilize the advantage of Coulomb gauge: the Coulomb propagator corrections
give the complete QCD $\beta$-function \cite{orsay,adler,robertson}, that permits
a simple implementation of renormalization group improved perturbation theory.
The Coulomb gauge allows an extension of the present approach to confining potentials
as one uses in phenomenological analysis \cite{eichten}.

The motivation for this study is to set up a kind of constituent gluon model
for glueball bound states, similar to constituent quark model.
In the previous study, one made a Bogoliubov-Valatin transformation 
to a quasiparticle basis, with a BCS vacuum containing a $q \bar{q}$ condensate
(pairing model) \cite{adler}. In the present work we apply 
the method of flow equations \cite{wegner} to transform to an effective Hamiltonian,
which is block-diagonal in particle number space and 
describes quasiparticle excitations \cite{gubweg}.

The proposed approach utilizes the method of flow equations in theory
of strong interactions with the confinement that is embedded. 
Using flow equations we scale down the (normal ordered) 
canonical QCD Hamiltonian from the bare cut-off scale $\Lambda\rightarrow\infty$ 
to some moderate scale $\Lambda_0$ where the perturbation theory breaks down.
Renormalization is performed by absorbing all
(in this order of perturbation theory) UV-divergences in counterterm operators.
This procedure recovers the perturbative renormalization for Hamiltonians \cite{gw93},
which aims to reduce the cut-off sensitivity of observables. 
Due to the principle of renormalization group invariance 
the physical gluon stays massless through the perturbative scaling.
QCD coupling constant, renormalized to the third order, starts to grow,
that stops asymtotic freedom at the scale $\Lambda\sim\Lambda_0$ 
and forbids to use perturbation theory further.
Strong coupling constant, $\alpha_s\sim 1$, does not reflect the whole complexity
of strong interactions at low energies.
We introduce confinement as a linear rising potential,
that sets the (hadronic) scale for the gluon mass.
In the renormalization group sense this ``spoils'' the theory:
there arises the massive gluon mode, which breaks gauge invariance.
In the present approach confinement makes it possible to run down flow equations
in the ``nonperturbative'' region from $\Lambda_0$ to the hadronic scale,
where the effective Hamiltonian has block-diagonal form.
The higher orders in iterrative procedure are suppressed
by the inverse power of (heavy) gluon mass. By applying flow equations
to block-diagonalize the ``confined QCD'' Hamiltonian 
one gets a closed chain of decoupled equations up to leading order,
which can be solved analytically. The block-diagonal effective Hamiltonian,
with fixed number of quasiparticles (constituent gluons and quarks),
provides a constituent desription for hadronic observables \cite{gsjc99}.
In the case of pure gluodynamics, one gets a constituent gluon model.

\section{Canonical Hamiltonian of gluodynamics in the Coulomb gauge}
\label{sec:2}

The starting point is the canonical QCD Hamiltonian (pure gluodynamics)
in the Coulomb gauge (${\bf\nabla}\cdot{\bf A}=0$) \cite{zwanziger}. 
Physical degrees of freedom are the transverse gauge fields
${\bf A}$ and their conjugate transverse momenta ${\bf\Pi}$.
The complete form of the canonical QCD Hamiltonian
can be found also in \cite{robertson}. There the potential source of difficulty comes from
the Faddeev-Popov determinant $\cal J={\rm det}[{\bf\nabla}\cdot{\bf D}]$ 
(normalized by ${\rm det}[{\bf\nabla}^2]=1$),
which arises when gauge fixing constraint equations for the physical fields are solved. 
We expand perturbatively the canonical QCD Hamiltonian in the bare coupling constant,
keeping the terms to the second order.
In leading order Faddeev-Popov determinant reduces to unity and simplifies the instantaneous term
to have the pure Coulomb behavior. 
The canonical Hamiltonian in the gluon sector takes the form
\begin{eqnarray}
{\rm H}_{\rm can} ={\rm H}_0+{\rm H}_{\rm int}
\,.\label{eq:2.1}\end{eqnarray}
The free gluon Hamiltonian, defined as $ {\rm H}_0 = {\rm H}_{\rm can}(g=0)$, 
is 
\begin{eqnarray}
{\rm H}_0 = {\rm Tr}\int d {\bf x} \left( {\bf\Pi}^2({\bf x})
+ {\bf B}^2_{A}({\bf x}) \right)
\,,\label{eq:2.2}\end{eqnarray}
which is written through the abelian component of magnetic field ${\bf B}_{A}$,
since the nonabelian magnetic field ${\bf B}={\bf B}^a T^a$ has components
$B_i^a=\epsilon_{ijk}\nabla_jA_k^a+\frac{g}{2}\epsilon_{ijk}f^{abc}A_j^bA_k^c$.
The trace in Eq. (\ref{eq:2.2}) is understood in color space.
The interaction part includes the (nonabelian) gluon terms: triple- and four-gluon vertices,
and the Coulomb instantaneous interaction
\begin{eqnarray}
{\rm H}_{\rm int} = {\rm Tr}\int d {\bf x} 
              \left( {\bf B}^2({\bf x})-{\bf B}^2_{A}({\bf x}) \right)
            + {\rm H}_C 
\,.\label{eq:2.2a}\end{eqnarray}
The Coulomb instantaneous interaction is given
\begin{eqnarray}
 {\rm H}_C =-\frac{1}{2}\int d{\bf x}d{\bf y}\rho^a({\bf x})V(|{\bf x}-{\bf y}|)\rho^a({\bf y})
\,,\label{eq:2.3}\end{eqnarray}
with the Coulomb potential $V(|{\bf x}-{\bf y}|)=-\alpha_s/|{\bf x}-{\bf y}|$
($\alpha_s=g^2/4\pi$), and color charge density (only gluon component) is
$\rho^a({\bf x})=f^{abc}{\bf A}^b({\bf x}){\bf\Pi}^c({\bf x})$.

We proceed in a standard way and express the Hamiltonian Eq. (\ref{eq:2.1})
in the Fock representation. The minimum ground state of the canonical Hamiltonian 
is achieved in the perturbative Fock space.
Therefore, we choose the trivial (perturbative) vacuum $|0\rangle$
and construct the perturbative basis of free (current) particles:
$a^{\dagger}({\bf k})|0\rangle$ creates one (perturbative) gluon with zero mass, etc.,
and the vacuum is defined as $a|0\rangle=0$.  
We decompose the gauge fields
\begin{eqnarray}
 && A_i^a({\bf x}) = \int\frac{d{\bf k}}{(2\pi)^3}\frac{1}{\sqrt{2\omega_{\bf k}}}
    [a_i^a({\bf k})+a_i^{a\dagger}(-{\bf k})]{\rm e}^{i{\bf k}{\bf x}} \nonumber\\
 && \Pi_i^a({\bf x}) = -i\int\frac{d{\bf k}}{(2\pi)^3}\sqrt{ \frac{\omega_{\bf k}}{2} }
    [a_i^a({\bf k})-a_i^{a\dagger}(-{\bf k})]{\rm e}^{i{\bf k}{\bf x}} 
\,,\label{eq:2.4}\end{eqnarray}
where the gluon energy $\omega_{\bf k}=|{\bf k}|$. 
The canononical commutation relation in momentum space is
\begin{eqnarray}
[a_i^a({\bf k}),a_j^{b\dagger}({\bf k}^{\prime})]=(2\pi)^3\delta^{ab}\delta^{(3)}
({\bf k}-{\bf k}^{\prime})D_{ij}({\bf k})
\,,\label{eq:2.5}\end{eqnarray}
where the operators $a_i^a$ are represented in terms of polarization vectors as \\
$a_i^a({\bf k})=\sum_{\lambda=1,2}\epsilon_i({\bf k},\lambda)a^a({\bf k},\lambda)$;
and the polarization sum is 
\begin{eqnarray}
D_{ij}({\bf k})=\sum_{\lambda=1,2}
\epsilon_i({\bf k},\lambda)\epsilon_j({\bf k},\lambda)
=\delta_{ij}-{\hat k}_i{\hat k}_j
\,,\label{eq:2.6}\end{eqnarray}
with the unit vector ${\hat k}_i=k_i/k$. The condition of transversality
is written 
\begin{eqnarray}
{\bf k}\cdot{\bf a}^a({\bf k})={\bf k}\cdot{\bf a}^{a\dagger}({\bf k})=0
\,,\label{eq:2.7}\end{eqnarray}
and hence $k_i\cdot D_{ij}({\bf k})=0$.

We represent the Hamiltonian in the perturbative Fock space, Eq. (\ref{eq:2.4}),
and perform normal ordering in the trivial vacuum state $|0\rangle$. 
In addition to the canonical terms, normal ordering
of the Coulomb interaction and the four-gluon vertex
gives rise to one-body and condensate operators.
 
{\bf The free gluon} part  Eq. (\ref{eq:2.2}) includes 
the kinetic term
\begin{eqnarray}
 H_0 &=& \frac{1}{2}\int\frac{d{\bf k}}{(2\pi)^3}
\left[(\frac{{\bf k}^2}{\omega_{\bf k}}+\omega_{\bf k})
a_i^{a\dagger}({\bf k})a_i^a({\bf k})
+(\frac{{\bf k}^2}{\omega_{\bf k}}-\omega_{\bf k})
\frac{1}{2}( a_i^a({\bf k})a_i^a(-{\bf k})+ {\rm h.c.} ) \right] \nonumber\\
&=& \int\frac{d{\bf k}}{(2\pi)^3}\omega_{\bf k} 
a_i^{a\dagger}({\bf k})a_i^a({\bf k}) 
+\frac{1}{2}\int\frac{d{\bf k}}{(2\pi)^3}
(\frac{{\bf k}^2}{\omega_{\bf k}}-\omega_{\bf k}) \nonumber\\
& \times &
(a_i^{a\dagger}({\bf k})a_i^a({\bf k})
+\frac{1}{2}( a_i^a({\bf k})a_i^a(-{\bf k})+ {\rm h.c.} ))
= \tilde H_0 + ( H_0 - \tilde H_0)
\,,\label{eq:2.8}\end{eqnarray}
and the condensate term from normal-ordering of Eq. (\ref{eq:2.8})
\begin{eqnarray}
 O_0 = \frac{1}{2}(N_c^2-1){\bf V}
\int\frac{d{\bf k}}{(2\pi)^3}(\frac{{\bf k}^2}{\omega_{\bf k}}+\omega_{\bf k})
\,,\label{eq:2.9}\end{eqnarray}
where the volume is ${\bf V}=(2\pi)^3\delta^{(3)}(0)$.
In the perturbative basis, $\omega_{\bf k}=|{\bf k}|$, 
the free Hamiltonian Eq. (\ref{eq:2.8}) reduces to the canonical part
\begin{eqnarray}
\tilde H_0 =\int\frac{d{\bf k}}{(2\pi)^3}\omega_{\bf k}
a_i^{a\dagger}({\bf k})a_i^a({\bf k}) 
\,,\label{eq:2.9a}\end{eqnarray}
that defines the Fock space: $\tilde H_0|n\rangle=E_n|n\rangle$, where
${|n\rangle}$ is a set of eigenstates and 
the energy of the state $|n\rangle$ is given by a sum of one-particle energies 
$E_n=\sum_m^n \omega_{{\bf k}_m}$.\\ 
{\bf The nonabelian gluon} part includes
in the order ${\cal O}(g)$ the triple-gluon coupling 
\begin{eqnarray}
H_g^{(1)} &=& \frac{ig}{2\sqrt{2}}f^{abc}
\int\left(\prod_{n=1}^{3} {d{\bf k}_n\over(2\pi)^3}\right)
(2\pi)^3\delta^{(3)}(\sum_m {\bf k}_m)
\frac{k_{1j}}{\sqrt{\omega_1\omega_2\omega_3}} \label{eq:2.10} \\
& & \times {\bf :}
\left[a^a_i({\bf k}_1) + {a^a_i}^\dagger(-{\bf k}_1)\right]
\left[a^b_j({\bf k}_2) + {a^b_j}^\dagger(-{\bf k}_2)\right]
\left[a^c_i({\bf k}_3) + {a^c_i}^\dagger(-{\bf k}_3)\right]
{\bf :}  \nonumber
\,,\end{eqnarray}
where in the following $\omega_1\equiv \omega_{{\bf k}_1}$, etc. \\
In the order ${\cal O}(g^2)$ the normal-ordered four-gluon vertex
\begin{eqnarray}
H_{g}^{(2)} & &  ={\alpha_s\pi\over4} f^{abc}f^{ade}
\int\left(\prod_{n=1}^{4} {d{\bf k}_n\over(2\pi)^3}\right)
(2\pi)^3\delta^{(3)}(\sum_m {\bf k}_m) 
{1\over\sqrt{\omega_1\omega_2\omega_3\omega_4}} \label{eq:2.11} \\
& &\hspace{-2cm}
\times {\bf :}
\left[ a^b_i({\bf k}_1) + {a^b_i}^\dagger(-{\bf k}_1)\right]
\left[ a^c_j({\bf k}_2) + {a^c_j}^\dagger(-{\bf k}_2)\right]
\left[ a^d_i({\bf k}_3) + {a^d_i}^\dagger(-{\bf k}_3)\right]
\left[ a^e_j({\bf k}_4) + {a^e_j}^\dagger(-{\bf k}_4)\right]
{\bf :} \nonumber
\,.\end{eqnarray}
Normal ordering in Eq. (\ref{eq:2.11}) produces one-body operator
\begin{eqnarray}
\Pi_g &=& \alpha_s\pi N_c
\int{d{\bf k}\thinspace d{\bf q}\over(2\pi)^6}
{1\over\omega_{\bf k}\omega_{\bf q}}[2\delta_{ij} - D_{ij}({\bf q})] 
\label{eq:2.12} \\
& &\quad\qquad\times\left[ {a^a_i}^\dagger({\bf k}) a^a_j({\bf k})
+\frac{1}{2}\left(a^a_i({\bf k}) a^a_j(-{\bf k})
+{\rm h.c.}\right)\right]  \nonumber
\,,\end{eqnarray}
and the condensate term
\begin{eqnarray}
{\cal O}_g &=& \frac{\alpha_s\pi}{4} N_c(N_c^2-1){\bf V}
\int{d{\bf k}\thinspace d{\bf q}\over(2\pi)^6}
{1\over\omega_{\bf k}\omega_{\bf q}}
\left(3-({\hat k}{\hat q})^2 \right)
\,.\label{eq:2.13}\end{eqnarray}
The {\bf Coulomb} piece Eq. (\ref{eq:2.3}) includes
the Coulomb instantaneous interaction
\begin{eqnarray} 
H_C^{(2)}& & = -{\alpha_s\over8} f^{abc}f^{ade} 
\int\left(\prod_{n=1}^{4} {d{\bf k}_n\over(2\pi)^3}\right)
(2\pi)^3\delta^{(3)}(\sum_m {\bf k}_m) 
\left({\omega_2\omega_4\over\omega_1\omega_3}\right)^{1/2}
\widetilde{V}({\bf k}_1+{\bf k}_2)  \label{eq:2.14} \\
& &\hspace{-2cm}
\times {\bf :}
\left[ a^b_i({\bf k}_1) + {a^b_i}^\dagger(-{\bf k}_1)\right]
\left[ a^c_i({\bf k}_2) - {a^c_i}^\dagger(-{\bf k}_2)\right]
\left[ a^d_j({\bf k}_3) + {a^d_j}^\dagger(-{\bf k}_3)\right]
\left[ a^e_j({\bf k}_4) - {a^e_j}^\dagger(-{\bf k}_4)\right]
{\bf :} \nonumber
\,,\end{eqnarray}
where $\widetilde{V}({\bf k})= 4\pi/{\bf k}^2$. The terms arising 
from normal ordering in Eq. (\ref{eq:2.14}) are \\  
the one-body operators
\begin{eqnarray} 
\Pi_C = {\alpha_s N_c\over4} & & \int
{d{\bf k}\thinspace d{\bf q} \over(2\pi)^6} \widetilde{V}({\bf k}+{\bf q})
\left({\omega_{\bf q}^2+\omega_{\bf k}^2 \over\omega_{\bf k}\omega_{\bf q}}\right)
D_{ij}({\bf q})
\left[ {a^a_i}^\dagger({\bf k}) a^a_j({\bf k})\right] \label{eq:2.15} \\
+ {\alpha_s N_c\over8} & & \int
{d{\bf k}\thinspace d{\bf q} \over(2\pi)^6} \widetilde{V}({\bf k}+{\bf q})
\left({\omega_{\bf q}^2 
- \omega_{\bf k}^2\over\omega_{\bf k}\omega_{\bf q}}\right)
D_{ij}({\bf q}) 
\left[a^a_i({\bf k}) a^a_j({\bf k}) + {\rm h.c.}\right]  \nonumber
\,,\end{eqnarray}
and the condensate term 
\begin{eqnarray}
{\cal O}_C &=& \frac{\alpha_s}{8}N_c(N_c^2-1){\bf V}
\int{d{\bf k}\thinspace d{\bf q}\over(2\pi)^6}
\widetilde{V}({\bf k}+{\bf q})
\left(\frac{\omega_{\bf q}}{\omega_{\bf k}}-1\right)
\left(1+({\hat k}{\hat q})^2\right)
\,.\label{eq:2.16}\end{eqnarray}
The one-body and condensate terms diverge in the ultraviolet region.

\section{Hamiltonian renormalization}
\label{sec:3}

As was mentioned in the previous section, normal-ordered canonical Hamiltonian
includes UV-divergent terms. The first thing to do is to regularize these divergences
(explicit form of regulator is specified below).
The cut-off sensitivity is removed by renormalization. Renormalization is performed
by adding counterterms, which are local operators with the symmetries of the canonical
Hamiltonian. In order to find explicit form of the counterterm operators 
we use flow equations, which evolve the Hamiltonian from the bare cut-off scale $\Lambda$ 
to some lower scale $\Lambda_0\ll\Lambda$. It is enough to find the gradient
of the Hamiltonian to define the counterterms. 
Also the form of the cutoff (regulating) function is specified by flow equations.   
 
In addition, new interactions, that do not change the number of particles,
are generated. The renormalized effective Hamiltonian is required to preserve
the strucure of the canonical Hamiltonian.

The renormalized Hamiltonian can be written in general   
\begin{eqnarray}
H_{\rm ren}(\Lambda)=H_{\rm can}+\delta X_{CT}(\Lambda)
\,,\label{eq:3.1}\end{eqnarray}
where $\delta X_{CT}(\Lambda)$ is a set of (unknown) counterterm operators.
The renormalized Hamiltonian depends implicitly on the cut-off $\Lambda$ 
through the counterterms
\footnote{Actually, the $\Lambda$-dependence arises for the renormalized Hamiltonian,
which is expressed in Fock space ($\Lambda$ regulates UV divergent loop integrals
in momentum space).
In the given (perturbative, $\omega_{\bf k}=|{\bf k}|$) basis
Eq. (\ref{eq:3.1}) reads
\begin{eqnarray}
{\bf :}H_{\rm ren}(\Lambda){\bf :}={\bf :}H_{\rm can}{\bf :}
+{\bf :}\delta X_{CT}(\Lambda){\bf :}
\,,\label{eq:3.1a}\end{eqnarray}
where ``{\bf :}'' stands for normal-ordering in the (perturbative) vacuum.
}. 

Primary aim of flow equations is to reduce the many-body problem of quantum
field theory to a few-body one.
The idea is to find a unitary transformation that transforms 
the Hamiltonian operator to a block-diagonal form, where each block conserves
the number of particles. 
Generally, the Hamiltonian matrix can be represented
in the particle number space as
\begin{eqnarray}
   H = \left(
   \begin{array}{cc}
      PHP & PHQ \\
      QHP & QHQ
   \end{array}\right)
\,,\label{eq:3.1b}\end{eqnarray}  
where $P$ and $Q$ are projection operators on the subspaces
with different particle number content.
The flow equations \cite{wegner} bring the Hamiltonian matrix Eq. (\ref{eq:3.1b})
to the form
\begin{eqnarray}
   H_{\rm eff} =\left( 
   \begin{array}{cc}
     PH_{\rm eff}P & 0            \\
     0             & QH_{\rm eff}Q
   \end{array}\right)
\,,\label{eq:3.1c}\end{eqnarray} 
where the two blocks of the effective Hamiltonian 
decouple from each other. It may be simpler then to solve 
for bound states within one block, say $PH_{eff}P$,
than to diagonalize the complete Hamiltonian, Eq. (\ref{eq:3.1b}),
of the original problem. Since, generally, the number of particles
in $P$ and $Q$ spaces is arbitrary, one can reduce in this way
the bound state problem with many particles
to a few-body problem.

The renormalization of possible ultraviolet divergences
can be done also by the method of flow equations. 
By using flow equations to block-diagonalize the Hamiltonian 
one eliminates the particle number changing contributions of the latter  
not in one step but rather continuous for the different energy differences 
in sequence. This procedure enables one to separate 
the ultraviolet divergent contributions and find
the counterterms associated with these divergences. 
This covers the UV-renormalization for Hamiltonians \cite{gw93}.

There is only one scale, introduced in the renormalized Hamiltonian 
by flow equations -- 
the cutoff scale $\Lambda$, which defines the size    
of the canonical ("bare") Hamiltonian matrix in the "energy" space,
and in the effective Hamiltonian plays the role of the regulator 
in the divergent loop integrals.
This can be compared with the similarity scheme \cite{gw93},
where the effective Hamiltonian is band-diagonal 
in the "energy" space and one introduces therefore two scales -- 
the $\Lambda$ which is the size of the Hamiltonian matrix 
and the $\lambda\ll\Lambda$ which is the width of the band,
and all the matrix elements of the effective Hamiltonian
are squeezed in a band, i.e. $|E_i-E_j|\le\lambda$.   
Generally, in the latter scheme not only the divergent contributions 
are regulated, but also any matrix element (its "external legs") 
is limited to permit the transition only between the states
with the energy differences less than $\lambda$.
One should not associate any physics with the second (axilary) scale $\lambda$,
and consider it as some technical tool  
to run the renormalization procedure
and find an explicit form of the counterterm operators.

The renormalized Hamiltonian, obtained by flow equations 
in the perturbative frame, preserves the form of
the (original) canonical Hamiltonian.
Therefore the renormalized Hamiltonian can be written
in terms of field variables, regardless of basis,
and can be decomposed in any nonperturbative Fock space,
provided canonical form.
Contrary to that, the effective Hamiltonian, obtained 
by the similarity scheme, can not be so trivially transformed
to the arbitrary (nonperturbative) basis, since the effective
Hamiltonian has the band-diagonal form only in the perturbative basis
and does not preserve it elsewhere.
This advantage of flow equations we exploit in section 4
to extend the calculations to the nonperturbative hadronic scales.
\\ 

The flow equations are written in differential form \cite{wegner} 
\begin{eqnarray}
&& \frac{dH(l)}{dl} =[\eta(l),H(l)] \nonumber\\
&& \eta(l) =[H_d(l),H_r(l)]
\,.\label{eq:3.2}\end{eqnarray}
where the Hamiltonian contains two pieces, $H(l)=H_d(l)+H_r(l)$,
with $H_d(l)$ including particle-number conserving (diagonal) 
and $H_r(l)$ - particle-number changing (rest) terms of Hamiltonian;
$\eta(l)$ is the generator of the unitary transformation
\footnote{
Unitary transformation reads
\begin{eqnarray}
H(l,l_0)=U^{-1}(l,l_0=0)H(l_0)U(l,l_0=0)
\,,\label{eq:3.2a}\end{eqnarray}
where 
\begin{eqnarray}
U(l,l_0=0)={\rm P exp}\left(\int_{l_0}^l \eta(l^{\prime})dl^{\prime}\right)
\,,\label{eq:3.2b}\end{eqnarray}
and $P$ stands for ordering along the flow parameter $l$.
The generator $\eta$ satisfies $\eta^{\dagger}=-\eta$.
};
$l$ is the flow parameter with the connection to the energy scale
$l=1/\lambda^2$, $l_0=0$ corresponds to the bare cut-off $\Lambda\rightarrow\infty$.
For the choice of the generator, given by Eq. (\ref{eq:3.2}), one ultimately 
eliminates the particle-number changing part of the Hamiltonian,
i.e. $H_r(l\rightarrow\infty)=0$ 
(actually the 'rest' part is exponentially suppressed for large $l$).
In this approach the generator includes only the terms 
from particle number changing sectors.

Decomposed in the perturbative basis ($\omega_{\bf k}=|{\bf k}|$) 
and normal ordered in the trivial vacuum $|0\rangle$, 
the canonical Hamiltonian, Eq. (\ref{eq:2.8})-Eq. (\ref{eq:2.16}),
is regularized by the bare cutoff
$\Lambda\rightarrow\infty$, i.e. the Hamiltonian is written
through the bare parameters
(bare couling constants and masses) at $l_0=1/\Lambda^2$.
The regulated canonical Hamiltonian is the starting point
for the further calculations in section 3.

Technically it is convenient to write flow equations
for the coefficients before the operator structures.
Flow equations are written selfconsistently 
for the coupling constants and 
masses (energies), that are the functions of the flow parameter $l$
and, generally, the functions of the momenta:
say, in Eq. (\ref{eq:2.8})-Eq. (\ref{eq:2.16}) to the leading order
the (triple-) coupling constant and the gluon energy become 
$g({\bf k}_1,{\bf k}_2,{\bf k}_3);l$ and $\omega({\bf k};l)$
as functions of in- and out- going momenta involved in the given sector.

We evolve the Hamiltonian with flow equations to the second order.
The energies get a correction in the order $O(g^2)$,
i.e. $\omega({\bf k};l)=|{\bf k}| + O(g^2)$.
Therefore the free Hamiltonian, Eq. (\ref{eq:2.8}),
is $H_0=\tilde H_0 + O(g^2)$,
with $\tilde H_0$ given by Eq. (\ref{eq:2.9a}).
To the leading order we include triple-gluon vertex 
into the ``rest'' part $H_r$ and other terms of canonical Hamiltonian
in the diagonal part $H_d$, i.e.
\begin{eqnarray}
H_r &=& H_g^{(1)} + O(g^2) \nonumber\\
H_d &=& \tilde H_0 + O(g^2)
\end{eqnarray}
where $O(g^2)$ stands for the second order terms.

The triple-gluon vertex, obtained by symmetrization of Eq. (\ref{eq:2.10}), 
reads 
\begin{eqnarray}
H_g^{(1)} &=& \frac{i}{2\sqrt{2}}f^{abc}
\int\left(\prod_{n=1}^{3} {d{\bf k}_n\over(2\pi)^3}\right)
(2\pi)^3\delta^{(3)}(\sum_m {\bf k}_m)
\frac{1}{\sqrt{\omega_1\omega_2\omega_3}} 
\Gamma_{ijk}({\bf k}_1,{\bf k}_2,{\bf k}_3)  \label{eq:3.4} \\
& &\hspace{-2cm}
\times {\bf :}
\left[g_0({\bf k}_1,{\bf k}_2,{\bf k}_3;l)
a^a_i({\bf k}_1) a^b_j({\bf k}_2) a^c_k({\bf k}_3) 
+3 g_1({\bf k}_1,{\bf k}_2,{\bf k}_3;l)
a^{a\dagger}_i(-{\bf k}_1) a^b_j({\bf k}_2) a^c_k({\bf k}_3)
+{\rm h.c.} \right]
{\bf :}  \nonumber
\,,\end{eqnarray}
with the form $\Gamma_{ijk}$ given by
\begin{eqnarray}
\Gamma_{ijk}({\bf k}_1,{\bf k}_2,{\bf k}_3)=
\frac{1}{6}\left[
(k_1-k_3)_j\delta_{ik}+(k_2-k_1)_k\delta_{ij}+(k_3-k_2)_i\delta_{jk} \right]
\,,\label{eq:3.5}\end{eqnarray}
which satisfies 
\begin{eqnarray}
&& \Gamma_{ijk}(-{\bf k}_1,-{\bf k}_2,-{\bf k}_3)=
(-1)\Gamma_{ijk}({\bf k}_1,{\bf k}_2,{\bf k}_3)
\nonumber\\
&& \Gamma_{jik}({\bf k}_2,{\bf k}_1,{\bf k}_3)=
(-1)\Gamma_{ijk}({\bf k}_1,{\bf k}_2,{\bf k}_3)
\,,\label{eq:3.6}\end{eqnarray}
In Eq. (\ref{eq:3.4}) $g_0(l)$ and $g_1(l)$ are coupling constants defined further.
To the leading order the generator of unitary transformation, 
defined as $\eta_g^{(1)}=[\tilde H_0,H_g^{(1)}]$, is given
\begin{eqnarray}
\eta_g^{(1)} &=& \frac{i}{2\sqrt{2}}f^{abc}
\int\left(\prod_{n=1}^{3} {d{\bf k}_n\over(2\pi)^3}\right)
(2\pi)^3\delta^{(3)}(\sum_m {\bf k}_m)
\frac{1}{\sqrt{\omega_1\omega_2\omega_3}} 
\Gamma_{ijk}({\bf k}_1,{\bf k}_2,{\bf k}_3) \label{eq:3.7} \\
& & \hspace{-2cm}
\times {\bf :}
\left[\eta_0({\bf k}_1,{\bf k}_2,{\bf k}_3;l)
a^a_i({\bf k}_1) a^b_j({\bf k}_2) a^c_k({\bf k}_3) 
+3 \eta_1({\bf k}_1,{\bf k}_2,{\bf k}_3;l)
a^{a\dagger}_i(-{\bf k}_1) a^b_j({\bf k}_2) a^c_k({\bf k}_3)
-{\rm h.c.} \right]
{\bf :}  \nonumber
\,,\label{eq:3.7a}\end{eqnarray}
where 
\begin{eqnarray}
\eta_0({\bf k}_1,{\bf k}_2,{\bf k}_3;l) &=&
D_0(\omega_1,\omega_2,\omega_3)g_0({\bf k}_1,{\bf k}_2,{\bf k}_3;l)
\nonumber\\
\eta_1({\bf k}_1,{\bf k}_2,{\bf k}_3;l) &=& 
D_1(\omega_1,\omega_2,\omega_3)g_1({\bf k}_1,{\bf k}_2,{\bf k}_3;l)
\,,\label{eq:3.8a}\end{eqnarray}
and the energy differences are
\begin{eqnarray}
D_0(\omega_1,\omega_2,\omega_3) &=& -(\omega_1+\omega_2+\omega_3)
\nonumber\\
D_1(\omega_1,\omega_2,\omega_3) &=& -(-\omega_1+\omega_2+\omega_3)
\,,\label{eq:3.8}\end{eqnarray}
In the order ${\cal O}(g)$ the flow equation reads
\begin{eqnarray}
\frac{dH_g^{(1)}}{dl}=[\eta_g^{(1)},\tilde H_0]
\,,\label{eq:3.9}\end{eqnarray}
that provides the solution for the coupling constants in Eq. (\ref{eq:3.4})   
\begin{eqnarray}
g_0(l)=g(0){\rm e}^{-D_0^2 l}\nonumber\\
g_1(l)=g(0){\rm e}^{-D_1^2 l}
\,,\label{eq:3.10}\end{eqnarray}
with $g(0)$ the bare coupling constant at $\Lambda\rightarrow\infty$;
we suppressed momentum idices in Eq. (\ref{eq:3.10}).
Equation (\ref{eq:3.10}) may be written in a general form \cite{gubpaulweg}
\begin{eqnarray}
g_i(l)=g(0)f(D_i;l)
\,,\label{eq:3.10a}\end{eqnarray}
where $f(D;l)$ is the similarity function with the properties
$f(D;l=0)=1$, $f(D;l\rightarrow\infty)=0$.
Such a more general form was first used by Glazek and Wilson \cite{gw93}.
In Eq. (\ref{eq:3.10}) the gaussian similarity function is used;
the sensitivity of bound state solution to the different choices of $f(D;l)$
was considered in \cite{gubpaulweg}.
Corresponding to Eq. (\ref{eq:3.10a}), the generator of the transformation is written  
\begin{eqnarray}
\eta_i(l) = -\frac{1}{D_i}\left(\frac{d{\rm ln}f(D_i;l)}{dl} \right)g_i(l)
\,,\label{eq:3.10b}\end{eqnarray}
Index $i$ shows, that coupling constant, $g_i$, and generator, $\eta_i$, 
are connected with the energy difference $D_i$.
We use this form, Eq. (\ref{eq:3.10a}) and Eq. (\ref{eq:3.10b}), in the calculations below.  

In the order ${\cal O}(g^2)$ the flow equation for the particle number 
conserving part, $H_d$, reads 
\begin{eqnarray}
\frac{dH_d^{(2)}}{dl}=[\eta_g^{(1)},H_g^{(1)}]
\,,\label{eq:3.11}\end{eqnarray}
that contribute to the two-body, one-body and condensate (v.e.v.) operators
in the Hamiltonian
\footnote{In the second order flow equation for 
the particle number changing terms, $H_r$, reads
\begin{eqnarray}
\frac{dH_r^{(2)}}{dl}=[\eta_g^{(1)},H_g^{(1)}]+[\eta_g^{(2)}, \tilde H_0]
\,,\label{eq:3.11a}\end{eqnarray}
where the second term with the generator $\eta_g^{(2)}=[\tilde H_0,H_r^{(2)}]$
makes $H_r^{(2)}$ to fall exponentially with flow parameter. 
}.
Further we consider all these sectors in sequence.

\subsection{Effective interaction (two-body sector)}

We calculate an effective gluon interaction in the color-singlet channel, that
describes the glueball bound state 
${\bf a}_q^{a\dagger}{\bf a}_{-q}^{a^\dagger}|0\rangle$.
For convenience we choose the c.m. system ${\bf P}=0$. An effective interaction is written
\begin{eqnarray}
V_{\rm eff} & &  = f^{abc}f^{a'b'c}
\int {d{\bf k} d{\bf q} \over(2\pi)^6}\frac{1}{\omega_{\bf k}\omega_{\bf q}} 
\widetilde{V}_{ii'jj'}({\bf k},{\bf q})
 \thinspace {\bf :}
{a^b_j}^\dagger({\bf q}){a^{b^\prime}_{j^\prime}}^\dagger(-{\bf q})
a^a_i({\bf k})a^{a^\prime}_{i^\prime}(-{\bf k})
{\bf :} 
\,.\label{eq:3.12}\end{eqnarray}
where the interaction kernel $\widetilde{V}_{ii'jj'}({\bf k},{\bf q})$ includes the terms
from the canonical Hamiltonian and new interactions, generated
to the second order by flow equations in the two-body sector Eq. (\ref{eq:3.11}).
Consider first the generated terms. 
The $t$-channel diagrams, arising from the commutator 
$[a_{1i}^aa_{2j}^{b\dagger}a_{3k}^{c\dagger},a_{1'i'}^{a'\dagger}a_{2'j'}^{b'}a_{3'k'}^{c'}]$
(diagram without backward motion) and from the commutator
$[a_{1i}^aa_{2j}^{b}a_{3k}^{c},a_{1'i'}^{a'\dagger}a_{2'j'}^{b'\dagger}a_{3'k'}^{c'\dagger}]$ 
(Z-graph in t-channel), do not contribute to the color singlet state;
only $s$-channel diagram, coming from the commutator
$[a_{1i}^{a\dagger}a_{2j}^{b}a_{3k}^{c},a_{1'i'}^{a'}a_{2'j'}^{b'\dagger}a_{3'k'}^{c'\dagger}]$,
has to be calculated (here the notation is $a_{1i}^{a}=a_{i}^{a}({\bf k}_1), etc.)$.
In the c.m. frame the flow equation for effective interaction is given 
\begin{eqnarray}
\frac{d V_{\rm gen}(l)}{dl}   &=& \frac{1}{8}f^{abc}f^{a'b'c}
\int {d{\bf k} d{\bf q} \over(2\pi)^6}
\frac{1}{\omega_{\bf k}\omega_{\bf q}\omega_{{\bf k}-{\bf q}}}
\nonumber\\
& \times &  
\Gamma_{ijk}({\bf k},-{\bf q},-({\bf k}-{\bf q}) )
\Gamma_{i'j'k'}({\bf k},-{\bf q},-({\bf k}-{\bf q}))
D_{kk'}({\bf k}-{\bf q})\cdot 4\cdot 9   \nonumber \\
& \times &  \left[
\eta_1({\bf q},{\bf k},{\bf k}-{\bf q};l)g_1({\bf k},{\bf q},{\bf k}-{\bf q};l) +
\eta_1({\bf k},{\bf q},{\bf k}-{\bf q};l)g_1({\bf q},{\bf k},{\bf k}-{\bf q};l)
\right]
\nonumber\\ 
& \times & {\bf :}
{a^b_j}^\dagger({\bf q}){a^{b'}_{j'}}^\dagger(-{\bf q})a^a_i({\bf k})a^{a'}_{i'}(-{\bf k})
{\bf :} 
\,,\label{eq:3.13}\end{eqnarray}
where factor $'4'$ is a number of possible permutations; and property for $\Gamma$-factors
Eq. (\ref{eq:3.6}) was used.
The sum of two terms $'\eta g'$
corresponds to the two different time-ordered diagrams in s-channel.
We introduce the factor $S_{ijk,i'j'k'}({\bf k},{\bf q})$ as
\begin{eqnarray}
\Gamma_{ijk}({\bf k},-{\bf q},-({\bf k}-{\bf q}))
\Gamma_{i'j'k'}({\bf k},-{\bf q},-({\bf k}-{\bf q})) =
\frac{1}{6}\frac{1}{6} 4 S_{ijk,i'j'k'}({\bf k},{\bf q})
\,.\label{eq:3.14}\end{eqnarray}
which is given due to transversality condition Eq. (\ref{eq:2.7}) 
\begin{eqnarray}
S_{ijk,i'j'k'}({\bf k},{\bf q})=(k_j\delta_{ik}-
\frac{1}{2}(k+q)_k\delta_{ij}+q_i\delta_{jk})
(k_{j'}\delta_{i'k'}-
\frac{1}{2}(k+q)_{k'}\delta_{i'j'}+q_{i'}\delta_{j'k'})
\,,\label{eq:3.15}\end{eqnarray}
this form is symmetric under transmutations ${\bf k}$ and ${\bf q}$.
The tensor structure of the effective interaction is defined by the contraction \\
$S_{ijk,i'j'k'}({\bf k},{\bf q})D_{kk'}({\bf k}-{\bf q})\equiv 
-V_{ij,i'j'}({\bf k},{\bf q})$,
that is given
\begin{eqnarray}
V_{ij,i'j'}({\bf k},{\bf q}) &=& -\left(\phantom{\frac{1}{2}}\hspace{-0.3cm}
q_iq_{i'}\delta_{jj'}+k_jk_{j'}\delta_{ii'}
+q_i(k_{j'}\delta_{i'j}-k_j\delta_{i'j'}) 
+q_{i'}(k_{j}\delta_{ij'}-k_{j'}\delta_{ij }) \right.\nonumber\\
&+& \left.\frac{k^2q^2}{({\bf k}-{\bf q})^2}
(1-({\hat k}{\hat q})^2)\delta_{ij}\delta_{i'j'}\right)
\nonumber\\
&=& -({\bf k}-{\bf q})^2 
\left( \phantom{\frac{1}{2}}\hspace{-0.3cm} n_in_{i'}\delta_{jj'}+n_jn_{j'}\delta_{ii'}
+n_in_j\delta_{i'j'}+n_{i'}n_{j'}\delta_{ij}
-n_in_{j'}\delta_{i'j}-n_{i'}n_j\delta_{ij'} \right.
\nonumber\\ 
&+& \left.\frac{k^2q^2}{({\bf k}-{\bf q})^4}(1-({\hat k}{\hat q})^2)
\delta_{ij}\delta_{i'j'}  \right)
\,,\label{eq:3.17}\end{eqnarray}
where $n_i=({\bf k}-{\bf q})_i/|{\bf k}-{\bf q}|$ is a unit vector.
Integrating flow equations Eq. (\ref{eq:3.13}) with the generator given by
Eq. (\ref{eq:3.10b}) 
provides the resulting generated interaction
$V_{\rm gen}(l\rightarrow\infty)\equiv V_{\rm gen}$
(the initial value is $V_{\rm gen}(l=0)=0$)
\begin{eqnarray}
V_{\rm gen} &=& -\alpha_s 2\pi f^{abc}f^{a'b'c}
\int {d{\bf k} d{\bf q} \over(2\pi)^6}
\frac{1}{\omega_{\bf k}\omega_{\bf q}\omega_{{\bf k}-{\bf q}} }
V_{ij,i'j'}({\bf k},{\bf q})
\left(\frac{\Theta(D_1,D_{1'})}{D_1}+\frac{\Theta(D_{1'},D_1)}{D_{1'}}\right)
\nonumber\\
& \times &
 {\bf :}
 {a^b_j}^\dagger({\bf q}){a^{b'}_{j'}}^\dagger(-{\bf q})a^a_i({\bf k})a^{a'}_{i'}(-{\bf k}) 
{\bf :}
\,,\label{eq:3.18}\end{eqnarray}
where we have introduced the $\Theta$-factor
\begin{eqnarray}
& & \Theta(D_1,D_2;l) = -\int_0^{l}\frac{df(D_1;l')}{dl'}f(D_2;l')dl' \nonumber\\
& & \Theta(D_1,D_2;l\rightarrow\infty) \equiv \Theta(D_1,D_2)
\,,\label{eq:3.19a}\end{eqnarray}
and the energy denominators are given
\begin{eqnarray}
& & D_1=-(-\omega_{\bf q}+\omega_{\bf k}+\omega_{{\bf k}-{\bf q}}) \nonumber\\
& & D_{1'}=-(-\omega_{\bf k}+\omega_{\bf q}+\omega_{{\bf k}-{\bf q}})
\,,\label{eq:3.19}\end{eqnarray}
the interaction kernel 
$V_{ij,i'j'}({\bf k},{\bf q})$ is defined in  Eq. (\ref{eq:3.17}).            
For the $\Theta$-factors the following holds
\begin{eqnarray}
& & \Theta(D,D;l) = \frac{1}{2}(1-f^2(D;l)) \nonumber\\
& & \Theta(D_1,D_2;l)+\Theta(D_2,D_1;l)=1-f(D_1;l)f(D_2;l) \nonumber\\
& & \frac{\Theta(D_1,D_2;l)}{D_1}+\frac{\Theta(D_2,D_1;l)}{D_2}
=\frac{1}{2}\left(\frac{1}{D_1}+\frac{1}{D_2}\right)(1-f(D_1;l)f(D_2;l))\nonumber\\ 
& &  +\frac{1}{2}\left(\frac{1}{D_1}-\frac{1}{D_2}\right)
(\Theta(D_1,D_2;l)-\Theta(D_2,D_1;l))
\,,\label{eq:3.20}\end{eqnarray}
that we exploit further. The first equation exibits the meaning of $\Theta$-factor:
it plays the role of the ultraviolet regulating function in divergent loop integrals
(see further).

We considered the gluon interactions, generated in the second order by flow equations,
that conserve the number of particles. The generator $\eta_g^{(1)}$ Eq. (\ref{eq:3.7})
produces to the second order
also the interactions, that change the number of particles,
for example interaction $W$ with the structure in second quantization
$a^{a\dagger}_{1i}a^{b}_{2j}a^{c}_{3k}a^{d}_{4l}$.
To eliminate these interactions we choose the generator of transformation 
$\eta=\eta_g^{(1)}+\eta_g^{(2)}$, where the second order generator is calculated by
$\eta_g^{(2)}=[H_0,W]$, that provides
the exponential damping of the particle number changing interaction $W$.
  
To maintain gauge invariance we add the normal-ordered Coulomb Eq. (\ref{eq:2.14}) and four-gluon
Eq. (\ref{eq:2.11}) terms of the canonical Hamiltonian in the color-singlet channel. 
Projecting these terms on the color singlet state in c.m. system gives
\begin{eqnarray}
V_C &=& -\frac{\alpha_s}{8}f^{abc}f^{a'b'c}
\int {d{\bf k} d{\bf q} \over(2\pi)^6}
\frac{(\omega_{\bf k}+\omega_{\bf q})^2}{\omega_{\bf k}\omega_{\bf q}}
\widetilde{V}({\bf k}-{\bf q}) \thinspace
\delta_{ij}\delta_{i'j'}
\thinspace {\bf :}
 {a^b_j}^\dagger({\bf q}){a^{b'}_{j'}}^\dagger(-{\bf q})a^a_i({\bf k})a^{a'}_{i'}(-{\bf k}) 
{\bf :} \nonumber\\
V_g &=& \frac{\alpha_s\pi}{4}
f^{abc}f^{a'b'c}
\int {d{\bf k} d{\bf q} \over(2\pi)^6}
\frac{2}{\omega_{\bf k}\omega_{\bf q}}
(\delta_{ii'}\delta_{jj'}-\delta_{ij'}\delta_{i'j})
\thinspace {\bf :}
 {a^b_j}^\dagger({\bf q}){a^{b'}_{j'}}^\dagger(-{\bf q})a^a_i({\bf k})a^{a'}_{i'}(-{\bf k}) 
{\bf :}
\,,\label{eq:3.21}\end{eqnarray}
Effective gluon interaction is given by a sum of terms
\begin{eqnarray}
V_{\rm eff} &=& V_{\rm gen}+V_C+V_g
\,,\label{eq:3.89}\end{eqnarray}
with generated interaction defined by Eq. (\ref{eq:3.18}).

Consider the leading behavior each of the terms Eq. (\ref{eq:3.89}). 
It is convenient to introduce
\begin{eqnarray}
 D &=& \frac{1}{2}(D_1+D_{1'})=-\omega_{{\bf k}-{\bf q}} \nonumber\\
 \delta D &=& \frac{1}{2}(D_1-D_{1'})=-(\omega_{\bf k}-\omega_{\bf q})
\,,\label{eq:3.20a}\end{eqnarray}
Actually the $\Theta$-factors in Eq. (\ref{eq:3.18}) are the functions
of the dimensionless parameter $\xi=\delta D/D$ \cite{gubpaulweg}.
Therefore the combination in Eq. (\ref{eq:3.18}) is written
\begin{eqnarray}
\frac{\Theta(D_1,D_{1'})}{D_1}+\frac{\Theta(D_{1'},D_1)}{D_{1'}}
=\frac{1}{D}\left(1+\frac{\xi^2-\xi\vartheta(\xi)}{1-\xi^2} \right)   
\,,\label{eq:3.20b}\end{eqnarray}
where $\vartheta(\xi)=\Theta(D_1,D_{1'})-\Theta(D_{1'},D_1)$, and
$\Theta(D_1,D_{1'})+\Theta(D_{1'},D_1)=1$.
Generated and normal-ordered canonical interactions contribute to
the effective interaction Eq. (\ref{eq:3.12}), respectively
\begin{eqnarray}
\widetilde{V}^{\rm gen}_{ii'jj'} &=& \frac{\alpha_s}{2}
\frac{4\pi}{\omega_{{\bf k}-{\bf q}}^2}
V_{ij,i'j'}({\bf k},{\bf q})
\left(1+\frac{\xi^2-\xi\vartheta(\xi)}{1-\xi^2}\right)
\nonumber\\
\widetilde{V}^{C}_{ii'jj'} &=& -\frac{\alpha_s}{8}\frac{4\pi}{({\bf k}-{\bf q})^2}
(\omega_{\bf k}+\omega_{\bf q})^2 \delta_{ij}\delta_{i'j'}
\nonumber\\
\widetilde{V}^{g}_{ii'jj'} &=& \frac{\alpha_s}{8}4\pi
(\delta_{ii'}\delta_{jj'}-\delta_{ij'}\delta_{i'j}) 
\,,\label{eq:3.20c}\end{eqnarray}
For real processes (when energy is conserved, $\xi=0$) 
the interaction terms take the form
\begin{eqnarray}  
\widetilde{V}^{\rm gen}_{ii'jj'} &=& 
\frac{\alpha_s}{2}\frac{4\pi}{\omega_{{\bf k}-{\bf q}}^2}
V_{ij,i'j'}({\bf k},{\bf q}) 
=\frac{\alpha_s}{2}\frac{4\pi}{({\bf k}-{\bf q})^2+m^2({\bf k}-{\bf q},\Lambda)}
V_{ij,i'j'}({\bf k},{\bf q})
\nonumber\\
\widetilde{V}^{C}_{ii'jj'} &=& -\frac{\alpha_s}{2}\frac{4\pi}{({\bf k}-{\bf q})^2}
\omega_{\bf k}^2 \delta_{ij}\delta_{i'j'} 
\,,\label{eq:3.20d}\end{eqnarray}
with phenomenological dispersion relation $\omega_{\bf k}^2={\bf k}^2+m^2({\bf k},\Lambda)$
where $m$ is dynamical gluon mass at the cut-off scale $\Lambda$, that we calculate further. 
The constituent gluon mass tends to zero as the scale is reduced and saturates
to some value as the scale is increased Fig. (\ref{fig.3}). 
One can speculate about the leading order behavior of the effective
interaction. The sum of Coulomb and generated terms behave as
\begin{eqnarray}  
\widetilde{V}_{\rm eff} \sim \frac{1}{({\bf k}-{\bf q})^2}
-\frac{1}{({\bf k}-{\bf q})^2+m^2({\bf k}-{\bf q},\Lambda)} 
\,,\label{eq:3.20d}\end{eqnarray} 
that has $1/k^2$ singularity for the large energy $\Lambda\gg 1GeV$ and $1/k^4$ singularity 
at low energy $\Lambda\le 1GeV$, corresponding to the Coulomb and the confining potentials, 
respectively.
This can be considered as a selfconsistent condition to include confinement 
in the calculations below.

\subsection{Polarization operator}

Diagrams with two contractions of the commutator $[\eta^{(1)},H^{(1)}]$
contribute to one-body operators 
\begin{eqnarray}
\Pi=\int\frac{d{\bf k}}{(2\pi)^3}\left( \Pi_{ij}^{ab}({\bf k})
a_{i}^{a\dagger}({\bf k})a_{j}^{b}({\bf k})
+M_{ij}^{ab}({\bf k})\thinspace
\frac{1}{2}\left( a_{i}^{a}({\bf k})a_{j}^{b}(-{\bf k})
           + {\rm h.c.} \right) \right)
\,,\label{eq:3.22}\end{eqnarray}
where polarization operators $\Pi_{ij}^{ab}({\bf k})$ are given by
the gluon loop integral
\begin{eqnarray}
\Pi_{ij}^{ab}({\bf k})=\int\frac{d{\bf q}}{(2\pi)^3}
\widetilde{\Pi}_{ij}^{ab}({\bf k},{\bf q})
\,,\label{eq:3.23}\end{eqnarray}
and analogous for $M_{ij}^{ab}({\bf k})$. Note, that while in general,
from rotational invariance, one-body operators 
$\Pi_{ij}^{ab}({\bf k})$ and $M_{ij}^{ab}({\bf k})$ have two components,
$\Pi_{ij}^{ab}({\bf k})=A^{ab}(k)\delta_{ij}+B^{ab}(k){\hat k}_i{\hat k}_j$
(the same for $M_{ij}^{ab}$), only the term proportional to $\delta_{ij}$
survive in the integral Eq. (\ref{eq:3.22}) due to transversality Eq. (\ref{eq:2.7}).
To separate $A(k)$ one can choose the vector $k=k_z$ and consider components
$i=j=x,y$. Then in Eq. (\ref{eq:3.23}), where $\widetilde{\Pi}_{ij}^{ab}({\bf k},{\bf q})$
has the terms of the type $\delta_{ij}$ and ${\hat q}_i{\hat q}_j$
(there are no terms proportional to ${\hat k}_i$ and ${\hat k}_j$),
one can substitute 
\begin{eqnarray}
{\hat q}_i{\hat q}_j\rightarrow\frac{1}{2}(1-({\hat k}{\hat q})^2)
\,.\label{eq:3.24}\end{eqnarray}
Considering only $\delta_{ij}$ component in Eq. (\ref{eq:3.23}) simplifies the following
calculations.

Making use of the second order flow equations Eq. (\ref{eq:3.11}), we have for
the gluon polarization operator
\\
\begin{eqnarray}    
& & \frac{d\Pi(l)}{dl} = \frac{1}{8}N_c\delta^{aa'}
\int {d{\bf k} d{\bf q} \over(2\pi)^6}
\frac{1}{\omega_{\bf k}\omega_{\bf q}\omega_{{\bf k}-{\bf q}}}
\nonumber\\
&\times&
\Gamma_{ijk}({\bf k},-{\bf q},-({\bf k}-{\bf q}))
\Gamma_{i'j'k'}({\bf k},-{\bf q},-({\bf k}-{\bf q}))
D_{jj'}({\bf q})D_{kk'}({\bf k}-{\bf q})\cdot 2\cdot 9\cdot 2
\nonumber\\
& & \left[ 
\left(\eta_1({\bf k},{\bf q},{\bf k}-{\bf q};l)g_1({\bf k},{\bf q},{\bf k}-{\bf q};l)
+\eta_0({\bf k},{\bf q},{\bf k}-{\bf q};l)g_0({\bf k},{\bf q},{\bf k}-{\bf q};l)\right)
\thinspace
a^{a\dagger}_i({\bf k})a^{a'}_{i'}({\bf k})
\right. \nonumber \\
&+& \left.
\left(\eta_1({\bf k},{\bf q},{\bf k}-{\bf q};l)g_0({\bf k},{\bf q},{\bf k}-{\bf q};l)
+\eta_0({\bf k},{\bf q},{\bf k}-{\bf q};l)g_1({\bf k},{\bf q},{\bf k}-{\bf q};l)\right)\right.
\label{eq:3.25}\\
&\times & \left. \frac{1}{2}(a^{a}_i({\bf k})a^{a'}_{i'}(-{\bf k})+{\rm h.c.})
\right] \nonumber 
\,,\end{eqnarray}
where the first term ($\eta_1g_1$) corresponds to a gluon loop without backward motion,
($\eta_0g_0$) stands for a $Z$-graph, and the last two terms ($\eta_1g_0$, $\eta_0g_1$)
are gluon loop diagrams with two incoming and two outgoing gluon lines;
by calculating Eq. (\ref{eq:3.25}) we have also used the property Eq. (\ref{eq:3.6}) 
for $\Gamma$-factors.

The tensor structure of polarization operator is given by a double contraction
of $S({\bf k},{\bf q})$ factor Eq. (\ref{eq:3.15}) with polarization sums 
in Eq. (\ref{eq:3.25}). We introduce therefore \\
$F_{ii'}({\bf k},{\bf q}) \equiv
S_{ijk,i'j'k'}({\bf k},{\bf q})D_{jj'}({\bf q})D_{kk'}({\bf k}-{\bf q})$,
where
\begin{eqnarray} 
F_{ii'}({\bf k},{\bf q}) &=&
\delta_{ii'}k^2q^2\left( (\frac{1}{q^2}+\frac{1}{({\bf k}-{\bf q})^2})
(1-({\hat k}{\hat q})^2)  \right) \nonumber\\
&+& 2q_iq_{i'}\left( 1-\frac{k^2}{2({\bf k}-{\bf q})^2}(1-({\hat k}{\hat q})^2) 
\right)
\,,\label{eq:3.26}\end{eqnarray}
Due to transversality condition we separate the $\delta_{ii'}$ component
in polarization operator, Eq. (\ref{eq:3.24}). The factor in Eq. (\ref{eq:3.26}) takes
the form  
\begin{eqnarray}
& & F_{ii'}({\bf k},{\bf q})\rightarrow \delta_{ii'}G({\bf k},{\bf q})/2 
\nonumber\\
& & G({\bf k},{\bf q})=2(1-({\hat k}{\hat q})^2)
\left( k^2+q^2+\frac{k^2q^2}{2({\bf k}-{\bf q})^2}
(1+({\hat k}{\hat q})^2) \right)
\,,\label{eq:3.27}\end{eqnarray}
Integrating flow equation (\ref{eq:3.25}) gives
\begin{eqnarray}
\delta\Pi &=& \Pi(l)-\Pi(l_0=0)
=  \alpha_s\pi N_c\int {d{\bf k} d{\bf q} \over(2\pi)^6}
\frac{1}{\omega_{\bf k}\omega_{\bf q}\omega_{{\bf k}-{\bf q}} }
G({\bf k},{\bf q})\nonumber\\
&\times& \left[
\left(\frac{\Theta(D_0,D_0;l)}{D_0}
+\frac{\Theta(D_1,D_1;l)}{D_1}\right)
a^{a\dagger}_i({\bf k})a^{a}_{i}({\bf k}) \right.
\nonumber\\  
&+& \left. \left(\frac{\Theta(D_0,D_1;l)}{D_0}
+\frac{\Theta(D_1,D_0;l)}{D_1}\right)
\frac{1}{2}(a^{a}_i({\bf k})a^{a}_{i}({-\bf k})+{\rm h.c.})
\right]
\,,\label{eq:3.28}\end{eqnarray}
where $\Theta$-factor is defined in Eq. (\ref{eq:3.19a}); the energy denominators
are given
\begin{eqnarray}
& & D_0=-(\omega_{\bf k}+\omega_{\bf q}+\omega_{{\bf k}-{\bf q}}) \nonumber\\
& & D_1=-(-\omega_{\bf k}+\omega_{\bf q}+\omega_{{\bf k}-{\bf q}})  
\,,\label{eq:3.29}\end{eqnarray}
Quantum corrections have the opposite sign than obtained by the flow equations,
since the flow equations carry out the scaling of Hamiltonian down
from high to low energies. Therefore, the polarization operator,
defined at some scale $l=1/\lambda^2$, is given
\begin{eqnarray}
& & \Pi_{\rm gen}(\lambda) = \frac{\alpha_s\pi}{2}N_c
\int {d{\bf k} d{\bf q} \over(2\pi)^6}
\frac{1}{\omega_{\bf k}\omega_{\bf q}\omega_{{\bf k}-{\bf q}}}
G({\bf k},{\bf q}) \nonumber\\
& &\hspace{1cm} \times \left[
\left(\frac{f^2(D_0;\lambda)}{D_0}+\frac{f^2(D_1;\lambda)}{D_1}\right)
a^{a\dagger}_i({\bf k})a^{a}_{i}({\bf k}) \right.
\nonumber\\ 
& &\hspace{1cm} +\left. \left( 
\left(\frac{1}{D_0}+\frac{1}{D_1}\right)f(D_0;\lambda)f(D_1;\lambda)
 + \left(\frac{1}{D_0}-\frac{1}{D_1}\right) \right.\right.
\label{eq:3.30} \\
& & \times \left.\left.
\left(\int^{\lambda}\frac{df(D_0;\lambda')}{d\lambda'}f(D_1;\lambda')d\lambda'
-\int^{\lambda}\frac{df(D_1;\lambda')}{d\lambda'}f(D_0;\lambda')d\lambda' \right)\right)
\frac{1}{2}(a^{a}_i({\bf k})a^{a}_{i}({-\bf k})+{\rm h.c.})
\right] \nonumber
\,,\end{eqnarray}
we have used identities Eq. (\ref{eq:3.20}) to simplify this expression.

In what follows we study the infrared behavior of the theory.
In this case, external momenta are soft and 
$D_0\sim D_1\sim -(\omega_{\bf q}+\omega_{{\bf k}-{\bf q}})\sim -2\omega_{\bf q}$ 
in Eq. (\ref{eq:3.30}). It is reasonable to assume for the regulating function
$f^2(D;\lambda)=f(D;\lambda)$, that corresponds to the rescaling of cut-off
$\lambda/\sqrt{2}\rightarrow\lambda$ and does not change the physical result.
The polarization operator, generated by the flow equations, 
takes the form
\begin{eqnarray}
\Pi_{\rm gen}(\lambda) &=& -\alpha_s\pi N_c
\int {d{\bf k} d{\bf q} \over(2\pi)^6}
\frac{G({\bf k},{\bf q})}{\omega_{\bf k}\omega_{\bf q}\omega_{{\bf k}-{\bf q}}} 
\frac{1}{\omega_{\bf q}+\omega_{{\bf k}-{\bf q}} }
{\rm e}^{-4q^2/\lambda^2}
\nonumber\\
& \times & 
\left( a^{a\dagger}_i({\bf k})a^{a}_{i}({\bf k}) 
+\frac{1}{2}(a^{a}_i({\bf k})a^{a}_{i}({-\bf k})+{\rm h.c.})
\right)  
\,.\label{eq:3.31}\end{eqnarray}
We have used the free dispersion relation $\omega_{\bf q}=|{\bf q}|$ in exponential 
factors $f(D;\lambda)$ up to the leading order. 
This is true for large cut-offs (see below). 
We add one-body operators, coming from normal-ordering of Coulomb
and four-gluon vertex, Eq. (\ref{eq:2.15}) and Eq. (\ref{eq:2.12}), respectively.
In these equations we also pick out the $\delta_{ij}$ component 
using Eq. (\ref{eq:3.24}).
The regulated polarization operators read
\begin{eqnarray}
\Pi_C(\lambda) &=& \frac{\alpha_s}{8}N_c\int {d{\bf k} d{\bf q} \over(2\pi)^6}
\frac{1}{\omega_{\bf k}\omega_{\bf q}}
\widetilde{V}({\bf k}-{\bf q})(1+({\hat k}{\hat q})^2){\rm e}^{-q^2/\lambda^2}
\nonumber\\
&\times & \left( (\omega_{\bf q}^2+\omega_{\bf k}^2)
a^{a\dagger}_i({\bf k})a^{a}_{i}({\bf k}) 
+(\omega_{\bf q}^2-\omega_{\bf k}^2)
\frac{1}{2}(a^{a}_i({\bf k})a^{a}_{i}({-\bf k})+{\rm h.c.}) \right)
\label{eq:3.32} \\
\Pi_g(\lambda) &=& \frac{\alpha_s\pi}{2}N_c\int {d{\bf k} d{\bf q} \over(2\pi)^6}
\frac{1}{\omega_{\bf k}\omega_{\bf q}}(3-({\hat k}{\hat q})^2)
{\rm e}^{-q^2/\lambda^2}
\nonumber\\
&\times & \left(a^{a\dagger}_i({\bf k})a^{a}_{i}({\bf k}) 
+\frac{1}{2}(a^{a}_i({\bf k})a^{a}_{i}({-\bf k})+{\rm h.c.}) \right) \nonumber
\,,\end{eqnarray}
We choose the regulating function in the loop integral over $d{\bf q}$ in Eq. (\ref{eq:3.32})
in accordance with the regulator in the generated term Eq. (\ref{eq:3.31})
to match the energy denominator. 
Note that for the generated term the regulator arises naturally
from the method of flow equations; for the normal-ordered
canonical terms one needs to introduce the regulator in the loop integral
in accordance with the generated terms.
The resulting polarization operator is given 
\begin{eqnarray}
\Pi(\lambda)=\Pi_{\rm gen}(\lambda)+\Pi_C(\lambda)+\Pi_g(\lambda)
\,,\label{eq:3.32a}\end{eqnarray}
with the terms defined in Eq. (\ref{eq:3.31}) and Eq. (\ref{eq:3.32}).
We use this expression in the next section 
to obtain gap equation.

At the end of this subsection we calculate the counterterm,
associated with the divergency of polarization operators.
The generated, Coulomb and four-gluon terms contribute
to the quadratically UV-divergent part, respectively
\begin{eqnarray}
{\rm div}\Pi(\Lambda) &=&
\frac{\alpha_s}{\pi}\frac{N_c}{3}\Lambda^2\left(-\frac{1}{4}+1+2 \right)
\nonumber \\
&\times&
\int {d{\bf k} \over(2\pi)^3} \frac{1}{2\omega_{\bf k}}
\left(a^{a\dagger}_i({\bf k})a^{a}_{i}({\bf k}) 
+\frac{1}{2}(a^{a}_i({\bf k})a^{a}_{i}({-\bf k})+{\rm h.c.}) \right)    
\,,\label{eq:3.33}\end{eqnarray}
The corresponding counterterm is defined therefore
\begin{eqnarray}
\delta X^{\prime}_{CT}(\Lambda) &=&  \int {d{\bf k} \over(2\pi)^3} 
\frac{m_{CT}^2}{2\omega_{\bf k}}
\left(a^{a\dagger}_i({\bf k})a^{a}_{i}({\bf k}) 
+\frac{1}{2}(a^{a}_i({\bf k})a^{a}_{i}({-\bf k})+{\rm h.c.}) \right)
\nonumber\\
&=& \frac{m_{CT}^2}{2}\int d{\bf x}
{\bf :}  A_i^a({\bf x})A_i^a({\bf x}) {\bf :}
\,,\label{eq:3.34}\end{eqnarray}
with the mass given
\begin{eqnarray}
m_{CT}^2= -\frac{\alpha_s}{\pi}N_c\frac{11}{12}\Lambda^2
\,,\label{eq:3.34a}\end{eqnarray}
When the quark sector is added using the same procedure,
the algebraic coefficient in the propagator correction reproduces 
the QCD $\beta$-function. This result was obtained
in the similarity scheme \cite{gw93} by D. Robertson et.al. \cite{robertson}.
This particular feature of the Coulomb gauge
provides the simple implementation of the renormalization group 
improved perturbation theory.

{\bf Gluon condensate}

The second order flow equation Eq. (\ref{eq:3.11}) for the condensate term reads
\begin{eqnarray}    
& & \frac{dO_{\rm gen}(l)}{dl} = \frac{1}{8}N_c(N_c^2-1){\bf V}
\int {d{\bf k} d{\bf q} \over(2\pi)^6}
\frac{1}{\omega_{\bf k}\omega_{\bf q}\omega_{{\bf k}-{\bf q}}}
\nonumber\\
& & \times
\Gamma_{ijk}({\bf k},-{\bf q},-({\bf k}-{\bf q}))
\Gamma_{i'j'k'}({\bf k},-{\bf q},-({\bf k}-{\bf q}))
D_{ii'}({\bf k})D_{jj'}({\bf q})D_{kk'}({\bf k}-{\bf q})\cdot 6 \cdot 2
\nonumber\\
& & \times
\eta_0({\bf k},{\bf q},{\bf k}-{\bf q};l)g_0({\bf k},{\bf q},{\bf k}-{\bf q};l)
\,,\label{eq:3.35}\end{eqnarray}
where the only commutator
$[a_{1i}^aa_{2j}^{b}a_{3k}^{c},a_{1'i'}^{a'\dagger}a_{2'j'}^{b'\dagger}a_{3'k'}^{c'\dagger}]$
contribute to the vacuum expectation value; factor $'6'$ is the number of possible permutations,
and ${\bf V}$ is the volume.

The tensor structure of the condensate term is given by the triple contraction
of the $S$-factor Eq. (\ref{eq:3.15}) with the polarization sums 
$S_{ijk,i'j'k'}({\bf k},{\bf q})D_{ii'}({\bf k})D_{jj'}({\bf q})D_{kk'}({\bf k}-{\bf q})\equiv 
G({\bf k},{\bf q})$, where function $G({\bf k},{\bf q})$ is given in Eq. (\ref{eq:3.27}).
Integrating flow equation (\ref{eq:3.35}) produces 
\begin{eqnarray}
O_{\rm gen}(l)-O_{\rm gen}(l_0=0) = 2\alpha_s\pi N_c(N_c^2-1){\bf V} 
\int {d{\bf k} d{\bf q} \over(2\pi)^6}
\frac{1}{\omega_{\bf k}\omega_{\bf q}\omega_{{\bf k}-{\bf q}} }
\frac{G({\bf k},{\bf q})}{3}\frac{\Theta(D_0,D_0;l)}{D_0}
\,.\label{eq:3.35a}\end{eqnarray}
The regulated condensate term is given 
\begin{eqnarray} 
O_{\rm gen}(\lambda) = \alpha_s\pi N_c(N_c^2-1){\bf V} 
\int {d{\bf k} d{\bf q} \over(2\pi)^6}
\frac{1}{\omega_{\bf k}\omega_{\bf q}\omega_{{\bf k}-{\bf q}} }
\frac{G({\bf k},{\bf q})}{3 D_0}
f^2(D_0;\lambda)  
\,,\label{eq:3.36}\end{eqnarray}
where $D_0$ is given in Eq. (\ref{eq:3.29}), and because of the scaling down
we changed the overall sign \footnote{
In the calculations below we use the symmetric form
\begin{eqnarray}
O_{\rm gen}(\lambda) &=& \alpha_s\pi N_c(N_c^2-1){\bf V} 
\int {d{\bf k}_1 d{\bf k}_2 d{\bf k}_3 \over(2\pi)^9}
\frac{1}{\omega_1\omega_2\omega_3 }
\nonumber\\
&\times & (2\pi)^3\delta^{(3)}({\bf k}_1+{\bf k}_2+{\bf k}_3) 
\frac{\widetilde{G}({\bf k}_1,{\bf k}_2,{\bf k}_3)}{3 D_0}
f^2(D_0;\lambda) 
\,,\label{eq:3.36a}\end{eqnarray}
where $D_0$ is given in Eq. (\ref{eq:3.8}), and  
\begin{eqnarray}
 \widetilde{G}({\bf k}_1,{\bf k}_2,{\bf k}_3) &=&
\left( [k_1^2+k_2^2+k_3^2]
[1-2 ({\hat k}_1{\hat k}_2) ({\hat k}_1{\hat k}_3) ({\hat k}_2{\hat k}_3) ]
 +k_1^2 ({\hat k}_1{\hat k}_2)^2({\hat k}_1{\hat k}_3)^2
\right.
   \label{eq:3.36b} \\
  &+& \left. k_2^2 ({\hat k}_1{\hat k}_2)^2({\hat k}_2{\hat k}_3)^2
+k_3^2 ({\hat k}_1{\hat k}_3)^2({\hat k}_2{\hat k}_3)^2
 \right) 
(2\pi)^3\delta^{(3)}({\bf k}_1+{\bf k}_2+{\bf k}_3) 
 \rightarrow G({\bf k},{\bf q}) \nonumber 
\,.\end{eqnarray} 
}.
We summarize the regulated generated and normal-ordered, 
Eq. (\ref{eq:2.16}) and Eq. (\ref{eq:2.13}), condensate terms 
\begin{eqnarray} 
O_{\rm gen}(\lambda) &=& -\alpha_s\pi N_c(N_c^2-1){\bf V} 
\int {d{\bf k} d{\bf q} \over(2\pi)^6}
\frac{1}{3\omega_{\bf k}\omega_{\bf q}\omega_{{\bf k}-{\bf q}} }
\frac{G({\bf k},{\bf q})}{(\omega_{\bf k}+\omega_{\bf q}+\omega_{{\bf k}-{\bf q}}) }
{\rm e}^{-(|{\bf q}|+|{\bf k}|+|{\bf k}-{\bf q}|)^2/\lambda^2}
\nonumber\\
O_C(\lambda) &=& \frac{\alpha_s}{8} N_c(N_c^2-1){\bf V}
\int{d{\bf k}\thinspace d{\bf q}\over(2\pi)^6}
\widetilde{V}({\bf k}-{\bf q})
\left(\frac{\omega_{\bf q}}{\omega_{\bf k}}-1\right)
\left(1+({\hat k}{\hat q})^2\right){\rm e}^{-(|{\bf q}|+|{\bf k}|)^2/\lambda^2}
\nonumber\\
O_g(\lambda) &=& \frac{\alpha_s\pi}{4} N_c(N_c^2-1){\bf V}
\int{d{\bf k}\thinspace d{\bf q}\over(2\pi)^6}
{1\over\omega_{\bf k}\omega_{\bf q}}
\left(3-({\hat k}{\hat q})^2 \right){\rm e}^{-({\bf q}|+|{\bf k}|)^2/\lambda^2}
\,,\label{eq:3.37}\end{eqnarray}
We have used the same reasoning as above to regularize the Coulomb 
and four-gluon terms. Radiative corrections to the gluon condensate,
Eq. (\ref{eq:2.9}), are summarized
\begin{eqnarray}
O(\lambda)=O_{\rm gen}(\lambda)+O_C(\lambda)+O_g(\lambda)  
\,.\label{eq:3.37sum}\end{eqnarray} 
Calculate the counterterm associated with ultraviolet divergent
condenstate part. The generated, Coulomb and four-gluon terms
contribute, respectively
\begin{eqnarray}
{\rm div}O (\Lambda) &=&
\frac{\alpha_s}{\pi}\frac{N_c(N_c^2-1){\bf V}}{3}\Lambda^2
\left(-\frac{1}{4}+1+2 \right)
\int {d{\bf k} \over(2\pi)^3} \frac{1}{2\omega_{\bf k}}
\,.\label{eq:3.37a}\end{eqnarray}
The counterterm is defined therefore
\begin{eqnarray}
\delta X^{\prime\prime}_{CT}(\Lambda) &=&  (N_c^2-1){\bf V}
\int {d{\bf k} \over(2\pi)^3} 
\frac{m_{CT}^2}{2\omega_{\bf k}}
= \frac{m_{CT}^2}{2}\int d{\bf x}
\langle 0| A_i^a({\bf x})A_i^a({\bf x})|0\rangle
\,,\label{eq:3.37b}\end{eqnarray}
where the mass given in Eq. (\ref{eq:3.34a}).
Combining the counteterms Eq. (\ref{eq:3.34}) and Eq. (\ref{eq:3.37b}),
one gets 
\begin{eqnarray}
\delta X_{CT} &=& \delta X^{\prime}_{CT}+\delta X^{\prime\prime}_{CT}
= m_{CT}^2 {\rm Tr}\int d{\bf x}{\bf A}^2({\bf x})
\,,\label{eq:3.37bc}\end{eqnarray}
which stands for the mass counterterm in the renormalized
to the second order Hamiltonian Eq. (\ref{eq:3.1}).
Note, that there is one and the same coefficient $m_{CT}$ for both counterterms
in one- and zero-body sectors. This suggests, that one can possible work
in field theoretical basis without decomposing the fields 
in Fock components (through creation and annihilation operators). 

At the end of this section we outline 
the Hamiltonian of gluodynamics, renormaized to the second order
\begin{eqnarray} 
{\rm H}_r = {\rm H}_{\rm can}+\delta X_{CT}
= {\rm H}_0+{\rm H}_{int}+\delta X 
= {\rm H}_0^r+{\rm H}_{int} 
\,,\label{eq:3.38}\end{eqnarray}
where ${\rm H}_{\rm can}$ is given in Eq. (\ref{eq:2.1}) and
the mass counterterm $\delta X_{CT}$  
Eq. (\ref{eq:3.37bc})
absorbes all possible UV-divergences, that appear to the second order.
The second order perturbative corrections to this Hamiltonian include 
the condensate terms
Eq. (\ref{eq:3.37}) and polarization operators Eq. (\ref{eq:3.32a}).
In one-body sector
\begin{eqnarray}
{H}_0^r &=& {H}_0 + \delta X' 
=\int\frac{d{\bf k}}{(2\pi)^3}\omega_{\bf k} 
a_i^{a\dagger}({\bf k})a_i^a({\bf k}) 
+\frac{1}{2}\int\frac{d{\bf k}}{(2\pi)^3}
(\frac{{\bf k}^2+m_{CT}^2}{\omega_{\bf k}}-\omega_{\bf k}) \nonumber\\
& \times &
(a_i^{a\dagger}({\bf k})a_i^a({\bf k})
+\frac{1}{2}( a_i^a({\bf k})a_i^a(-{\bf k})+ {\rm h.c.} ))   
\,,\label{eq:3.38a}\end{eqnarray}
where $m_{CT}^2$ is the mass counterterm Eq. (\ref{eq:3.34a}).
In the 'leading log' approximation (for the terms propotional to $\Lambda^2$)
the renormalized free Hamiltonian preserves the structure of the canonical
free part Eq. (\ref{eq:2.9a}), 
provided the second order radiative corrections $O(g^2)$ are included, i.e.
\begin{eqnarray}
H_0^r + O(g^2) = \tilde H_0
\,,\label{eq:3.38b}\end{eqnarray}
that manifests renormalization group invariance (RGI),
or coupling coherence in the context of Hamiltonian renormalization \cite{pw}.
The same holds in zero-body sector for the condensate operator
\footnote{
The renormalized condensate term is given
\begin{eqnarray}
O_0 &=& \frac{1}{2}(N_c^2-1){\bf V}
\int\frac{d{\bf k}}{(2\pi)^3}(\frac{{\bf k}^2+m_{CT}^2}
{\omega_{\bf k}}+\omega_{\bf k})  \nonumber\\
&=& (N_c^2-1){\bf V}
\int\frac{d{\bf k}}{(2\pi)^3}\omega_{\bf k}
+\frac{1}{2}(N_c^2-1){\bf V}
(\frac{{\bf k}^2+m_{CT}^2}
{\omega_{\bf k}}-\omega_{\bf k})   
\,,\label{eq:3.38c}\end{eqnarray}
where the second term vanishes (in the order $\Lambda^2$) 
when perturbative corrections are calculated.
}.

One can proceed with flow equations order by order to find 
(all) the counterterms systematically.  
RGI insures, that
the renormalized Hamiltonian preserves the form
of the (original) canonical Hamiltonian, but only the coupling constants 
and the mass operators (that are usually classified as relevant and 
marginal operators in renormalization group sense)
start to run with the cutoff scale.
(We do not consider here, at least in the few lowest orders of PT, 
possible irrelevant operators, that may cause new type of divergences 
than are carried by coupling constants and masses).

Using flow equations we run the renormalized Hamiltonian
in perturbative basis, Eq. (\ref{eq:3.1a}),
downwards from the bare cutoff $\Lambda$ to some intermediate 
scale $\Lambda_0\sim\Lambda_{QCD}$, where the perturbation theory 
breaks down. Due to the RGI the "physical gluon" stays massless 
through this perturbative scaling. We can not proceed 
with flow equations perturbatively further.
The result of this stage is the renormalized to the second order, 
effective Hamiltonian, defined at some compositness scale, 
${\bf :}H_{ren}(\Lambda,\Lambda_0){\bf :}$ with $\Lambda_0\sim\Lambda_{QCD}$
and $\Lambda\rightarrow\infty$ is UV bare cutoff,  
and semicolon means normal-ordering in the trivial vacuum $|0\rangle$.
Though the renormalized Hamiltonian is obtained in the perturbative
frame, it can be represented (regardless of the Fock basis) 
in terms of the fields ${\bf A}$ and ${\bf \Pi}$, Eq. (\ref{eq:3.38}).
It is a consequence of the RGI.
We denote the resulting renormalized Hamiltonian at the scale $\Lambda_0$
as $H_{ren}(\Lambda,\Lambda_0)$.

\section{Flow equations in the confining background}
\label{sec.:4}

We introduce confinement as a linear rising potential, that enables to run 
flow equations "nonperturbatively" until complete diagonalization of the Hamiltonian.
In the renormalization group sense this "spoils" the theory: 
there arises the massive gluon mode, which breaks the gauge invariance. 
In the present approach 
the confinement makes possible to bring by flow equations
the QCD Hamiltonian to a block-diagonal form 
with a fixed number of quasiparticles in each sector. 
The elementary degrees of freedom (quasiparticles) 
become constituent gluons (quarks), which acquire masses of order $GeV$.
Confinement (string tension) sets the (hadron) scale for the gluon mass.
The effective (block-diagonal) Hamiltonian, written in terms of  
massive gluon modes, provides a kind of the constituent gluon picture.
In the case of zero gluon masses one can not find 'sector' representation
for the effective Hamiltonian, because of uncontrolable creation
and annihilation of particles in vacuum. 

The instantaneous interaction, Eq. (\ref{eq:2.3}), containes two pieces, 
the sum of the Coulomb and confining potentials
\begin{eqnarray}
V(r) =-C_{adj}\frac{\alpha_s}{|{\bf r}|} + \sigma |{\bf r}|
\,,\label{eq:4.1}\end{eqnarray}
in configuration space. The Casemir operator in adjoint representation
is $C_{adj}=(N_c^2-1)/2N_c$, and  
$\sigma$ is string tension.
Denote the renormalized effective Hamiltonian with 
the confinement embeded, Eq. (\ref{eq:3.38}) and Eq. (\ref{eq:4.1}), 
as $H_{eff}(\Lambda,\Lambda_0)$.

As far as confinement is introduced the trivial vacuum $|0\rangle$  
and the perturbative basis of free (current) particles,
$\omega_{\bf k}=|{\bf k}|$, define no longer the "minimum" ground state.
Therefore, we introduce the (arbitrary) basis,    
where the gluon energy $\omega_{\bf k}$ is kept unknown,
and is defined further variationally.
Correspondingly, the (nontrivial) QCD vacuum $|\Omega\rangle$ 
is defined as $\alpha|\Omega\rangle =0$, and the Fock space 
of {\it constituent} particles is given:
$\alpha^{\dagger}|\Omega\rangle$
creates the quasiparticle with the energy $\omega_{\bf k}$, etc.
\footnote{ 
The change of basis  
from the (perturbative) current, $\omega_{\bf k}=|{\bf k}|$, 
to the (nonperturbative) constituent, with some $\omega_{\bf k}$, 
can be written as Bogoluibov-Valatin (BV) transformation  
from the "old", $a,a^{\dagger}$, to the "new", $\alpha,\alpha^{\dagger}$, 
operators 
\begin{eqnarray}
a_{\bf k}={\rm ch}\phi_{\bf k}\alpha_{\bf k}+
{\rm sh}\phi_{\bf k}\alpha_{-{\bf k}}^{\dagger}
\,,\label{eq:4.1a}\end{eqnarray} 
with BV angle $\phi_{\bf k}$ given by
${\rm ch}\phi_{\bf k}=1/2(\sqrt{k/\omega_{\bf k}}+
\sqrt{\omega_{\bf k}/k})$ 
The connection between 
the "old", $|0\rangle$, and the "new", $|\Omega\rangle$, vacuum states 
is given  
\begin{eqnarray}
|\Omega\rangle={\rm exp}\left( 
\frac{1}{2}\sum_k{\rm th}\phi_{\bf k} 
a^{\dagger}_{\bf k}a^{\dagger}_{-{\bf k}} \right) |0\rangle
\,.\label{eq:4.1c}\end{eqnarray}
It was used in the work \cite{ssjc96} to transform the QCD Hamiltonian
into the constituent basis.}
 
The renormalized effective Hamiltonian $H_{eff}(\Lambda,\Lambda_0)$
at the scale $\Lambda_0$, written through the physical fields
${\bf A}$ and ${\bf \Pi}$ and having confinement, 
is decomposed in the trial (constituent) basis and normal-ordered
with respect to QCD vacuum $|\Omega\rangle$. 
The resulting effective Hamiltonian is given 
by Eq. (\ref{eq:2.10})-Eq. (\ref{eq:2.16})
and Eq. (\ref{eq:3.38a}), Eq. (\ref{eq:3.38c}), where 
the unknown gluon energy $\omega_{\bf k}$
is variational parameter in the calculations.
We denote the effective Hamiltonian 
in constituent basis as ${\bf ::}H_{eff}(\Lambda_0){\bf ::}$,
where $''{\bf ::}''$ stands for normal-ordering in the QCD vacuum.

We combine the terms in the effective Hamiltonian 
in each particle number sector according
to the power of coupling constant $O(g^n)$ ($n=0,1,2$).
The higher order terms in the effective Hamiltonian
are suppressed by the inverse powers of (heavy) gluon mass,
which is of the order of hadron scale (see calculations below).
Variational calculations below give for the gluon energy
$\omega_{\bf k}=|{\bf k}| + m({\bf k})$ 
with the effective mass $m({\bf k})$,
where for small momenta $m({\bf k})$ tends to the value $\sim {\rm GeV}$ 
and vanishes
at high momenta.
Typically, in the second order of perturbation theory we have
\begin{eqnarray}
\alpha_s \frac{V_{12}}{(E_1-E_2)}
\,,\label{eq:small}\end{eqnarray}
where the matrix element of the canonical interaction $V_{12}$ is
of order of the inverse Bohr radius, $\sim {\rm MeV}$, and 
the energy denominator, 
given by combinations of gluon energies, is $\sim {\rm GeV}$.
Though for small momenta the coupling constant is not small,
$\alpha_s\sim 1$,
we have the small parameter $\sim 0.1-0.01$, Eq. (\ref{eq:small}),
due to the (heavy) effective gluon mass $m({\bf k}=0)$. 

In the absence of confinement the effective Hamiltonian 
preserves the form of canonical Hamiltonian due to RGI,
with the change $|{\bf k}|\rightarrow\omega_{\bf k}$ properly.
In the presence of confinement the canonical form is violated by 
the second order terms in the free Hamiltonian, 
Eq. (\ref{eq:3.38a}),  
which contribute higher orders $O(g^3)$, etc. in flow equations.
We aim to find the effective Hamiltonian
after the scaling downwards from $\Lambda_0$
to a hadron scale, say $\sqrt{\sigma}$.
Since the effective Hamiltonian preserves
the canonical form at least to the second order,
the "perturbative" terms obtained by flow equations in 
section 3 match the "nonperturbative" terms arising 
when applied flow equations to $H_{eff}(\Lambda,\Lambda_0)$.
Generally, this approach allows to include perturbative QCD corrections
into nonperturbative calculations of many-body techniques.

We run flow equations in the confining background to block-diagonalize 
the effective Hamiltonian ${\bf ::}H_{eff}(\Lambda,\Lambda_0){\bf ::}$ in 
the nonperturbative basis and to find consistently 
all the terms to the second order.   
Free Hamitlonian and the confining interaction are included in "diagonal" sector,
the triple-gluon vertex forms "nondiagonal" sectors, that should be eliminated.
The leading order generator of unitary transformation
is given by Eq. (\ref{eq:3.7a}) with the trial gluon energy $\omega_{\bf k}$.
It is important, that the zero order approximation in flow equations include
all particle number conserving terms of Hamiltonian.
In the previous work \cite{gw93} free Hamiltonian was chosen to start
iterative procedure. The better convergence to a bound state Hamiltonian
is achieved in the case of flow equations.
This enables to use an ``effective'' (trial) basis in calculations.
At low energies this basis tend to constituent degrees of freedom.
This description is close to the approach of Orsay group \cite{orsay}.
 
We bring the Hamiltonian ${\bf ::}H_{eff}(\Lambda,\Lambda_0){\bf ::}$ 
to a block-diagonal form, 
where diagonal blocks decouple from each other including the second order.
All calculations are the same as carried by renormalization procedure in section 3,
except for the change of the perturbative basis to the unknown nonperturbative basis,
fixed below (section 5) variationally.
Since block-diagonalization is performed,
the particle number changing interactions are eliminated (to the second order) completely. 

We outline the resulting block-diagonal effective Hamiltonian,
renormalized to the second order, denoted as $H_{eff}$
\begin{eqnarray} 
{\rm H}_{\rm eff} = \widetilde{\rm H}_{\rm can}+{\rm V}_{\rm gen}+\delta X_{CT}
\,,\label{eq:3.38}\end{eqnarray}
where $\widetilde{\rm H}_{\rm can}$ coicide with canonical Hamiltonian Eq. (\ref{eq:2.1})
except for the particle number changing terms, which are eliminated in the leading order
by the flow equations;
instead the new gluon interaction ${\rm V}_{\rm gen}$ is generated 
in two-body sector Eq. (\ref{eq:3.18}); the mass counterterm $\delta X_{CT}$ 
Eq. (\ref{eq:3.37bc})
absorbes all possible UV-divergences, that appear to the second order.
The second order perturbative corrections to this Hamiltonian include the condensate terms
Eq. (\ref{eq:3.37}) and polarization operators Eq. (\ref{eq:3.32a}).
  
Formally, the matrix elements of the effective Hamiltonian can be written
\begin{eqnarray}
& & \langle 0|H_{\rm eff}|0\rangle =
 O_0 + O_{\rm gen}(\Lambda) + O_C(\Lambda) + O_g(\Lambda)+\delta X^{\prime\prime}_{CT}
\nonumber\\
& & \langle 1|H_{\rm eff}|1\rangle = 
 H_0+\Pi_{\rm gen}(\Lambda)+\Pi_C(\Lambda)+\Pi_g(\Lambda)+\delta X^{\prime}_{CT}
\nonumber\\
& & \langle 2|H_{\rm eff}|2\rangle =
V_{\rm gen}(\lambda\rightarrow 0)+V_C(\lambda\rightarrow 0)+V_g(\lambda\rightarrow 0)
\,,\label{eq:3.37d}\end{eqnarray}
where the term $O_0$, $H_0$ are defined 
in Eq. (\ref{eq:2.9}) and Eq. (\ref{eq:2.8}),
respectively, the second order radiative corrections: condensate and polarization terms 
are in Eq. (\ref{eq:3.37sum}) 
and Eq. (\ref{eq:3.32a}), respectively,
the countertems are in Eq. (\ref{eq:3.37b}) and Eq. (\ref{eq:3.34}),
the effective color-singlet interaction is in Eq. (\ref{eq:3.89}).
Other possible two-body interactions, that change the particle number,
are eliminated by flow equations to the second order 
and give nontrivial contributions to the higher orders. 
In the next section we use Eq. (\ref{eq:3.37d}) to derive gap equation and Tamm-Dancoff
bound state equation.

\section{Glueball spectrum}
\label{sec.:3}
 
The result of the previous section is the renormalized 
effective Hamiltonian of gluodynamics up to the second order, Eq. (\ref{eq:3.38}). 
Equations for physical observables
based on the renormalized Hamiltonian are protected from UV-divergences.
In this section we derive gap equation for the constituent gluon mass.
Gluon dispersion relation and the effective gluon interaction
are used to solve for the glueball bound state within the Tamm-Dancoff approach.

\subsection{Gap equation}

We use the traditional way to derive gap equation.
In Bogoliubov-Valatin approach the trial vacuum state $|\Omega\rangle$
is optimized variationally, and the variational parameter is 
the Bogoliubov-Valatin angle of transformation.  
In our case the effective Hamiltonian has been transformed
to a block-diagonal form in a quasiparticle basis.
Hence the vacuum expectation value of the effective Hamiltonian
should be optimized in this vacuum; the variational parameter
is an unknown one-particle dispersion relation.
One has therefore
\begin{eqnarray}  
\frac{\delta \langle 0|H_{\rm eff}|0\rangle}{\delta\omega_{\bf k}}=0
\,,\label{eq:4.1}\end{eqnarray} 
that provides an equation for $\omega_{\bf k}$.
Making use of Eq. (\ref{eq:3.37d}) for the condensate terms
the following gap equation is obtained
\begin{eqnarray}
\omega_{\bf k}^2 &=& k^2+m_{CT}^2
\nonumber\\
&+& \frac{\alpha_s}{4} N_c
\int{d{\bf q}\over(2\pi)^3}{1\over\omega_{\bf q}}
\widetilde{V}({\bf k}-{\bf q})\left(1+({\hat k}{\hat q})^2 \right)
(\omega_{\bf q}^2-\omega_{\bf k}^2){\rm e}^{-q^2/\Lambda^2}
\nonumber\\
&+& \alpha_s\pi N_c
\int{d{\bf q}\over(2\pi)^3}{1\over\omega_{\bf q}}
\left(3-({\hat k}{\hat q})^2 \right)
{\rm e}^{-q^2/\Lambda^2}
\nonumber\\
&-& 2\alpha_s\pi N_c
\int{d{\bf q}\over(2\pi)^3}{1\over\omega_{\bf q}\omega_{{\bf k}-{\bf q}} }
\left(-\frac{1}{D_0}+\frac{\omega_{\bf k}}{D_0^2}\right)
G({\bf k},{\bf q}){\rm e}^{-4q^2/\Lambda^2}
\,,\label{eq:4.2}\end{eqnarray}
where $D_0=-(\omega_{\bf k}+\omega_{\bf q}+\omega_{{\bf k}-{\bf q}})$
and the mass counterterm $m_{CT}$ is defined in Eq. (\ref{eq:3.34a}).
To get Eq. (\ref{eq:4.2}) we have used for convenience the symmetrized 
form for the generated term, Eq. (\ref{eq:3.36a}).
By performing variation over external gluon energy the regulator in the condensate terms
reduces to the regulator, obtained in one-body sector
for polarization terms when external momentum is soft, i.e. $|{\bf k}|\ll |{\bf q}|$. 

The integral kernel of gap equation contains three terms. The most essential
is the first one (with $\widetilde{V}$), which forms the basis
of the given above approach. We solve Eq. (\ref{eq:4.2}) numerically
taking into account only the first term of the integral kernel and the corresponding
to it Coulomb mass counterterm
\begin{eqnarray} 
\widetilde{m}_{CT}^2 = -\frac{\alpha_s}{\pi}\frac{N_c}{3}\Lambda^2
\,.\label{eq:4.2ct}\end{eqnarray}
Potential $\widetilde{V}$ is presented as a sum of Coulomb
and confining potentials, Eq. (\ref{eq:4.1}), with the Fourier transform
\begin{eqnarray} 
\widetilde{V}({\bf k})=C_{adj}\alpha_s\frac{4\pi}{{\bf k}^2}+
\sigma \frac{1}{{\bf k}^4} 
\,,\label{eq:4.2int}\end{eqnarray}
where the latter interaction is added to obtain nonperturbative properties. 
The canonical Coulomb term (and also the neglected integral terms)
leads to the leading $\Lambda^2$ and the subleading $\ln\Lambda$
(corresponding to relevant and marginal operators in the context
of renormalization group)
UV-divergencies. These divergencies are regulated by the cut-off function
$R=exp(-q^2/\Lambda^2)$. Since the gap equation is renormalized 
to the second order, the mass counterterm cancels the leading UV-divergency.
Note, that confining potential also contribute a term of the type $\ln\Lambda$
in UV region. The logarithmic cut-off dependence of the Coulomb
and the confining integral terms leads to a slow logarithmic growth
of the mass gap with $\Lambda$. This behavior can be absorbed by the running
of the strong coupling constant $\alpha_s$.

\begin{figure}[!htb]
\begin{center}
\input{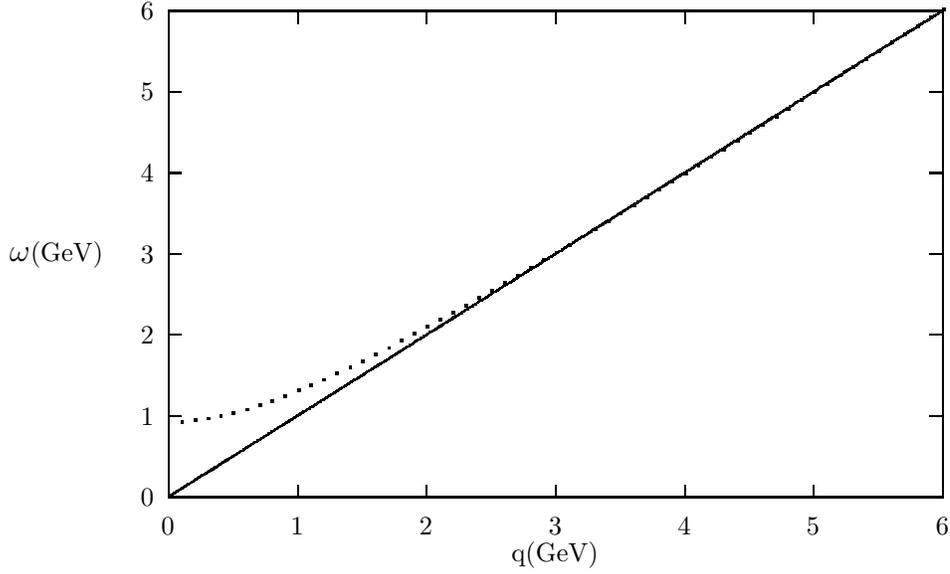}
\end{center}
\caption{One particle dispersion relation.
Dots represent the numerical solution 
of gap equation $\omega({\bf k})$ 
($\alpha_s=0.4, \sigma=0.18 GeV^2, \Lambda=4 GeV, N_c=3$), 
the solid line stays for the free
dispersion relation $\omega({\bf k})=k$.}
\label{fig.1}
\end{figure}

\begin{figure}[!htb]
\begin{center}
\input{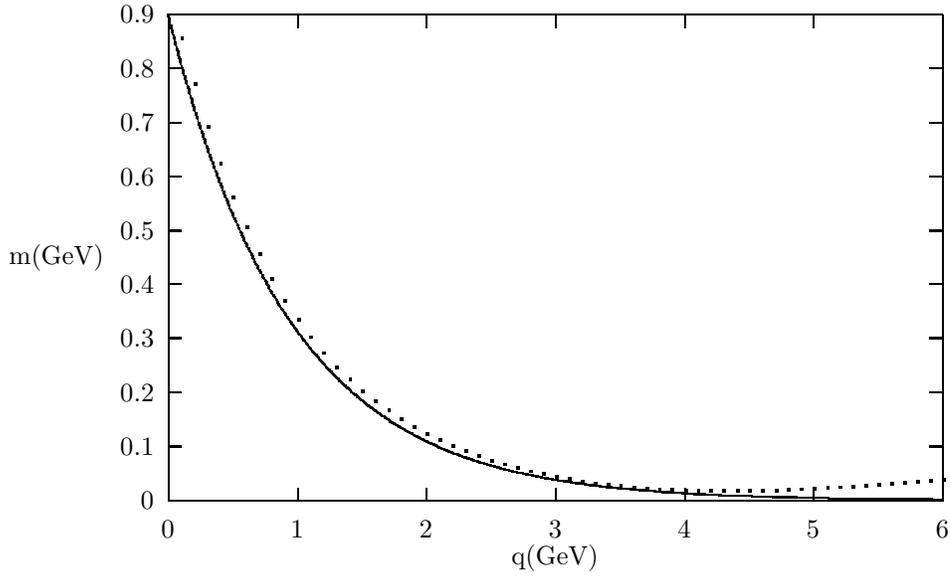}
\end{center}
\caption{Gluon mass. Dots represent the numerical solution 
for $m({\bf k})=\omega({\bf k})-k$
($\alpha_s=0.4, \sigma=0.18 GeV^2, \Lambda=4 GeV, N_c=3$), 
the solid line is a fit
$m({\bf k})=0.9*{\rm exp}(-k/0.95)$ (parameters are in GeV).}
\label{fig.2}
\end{figure}

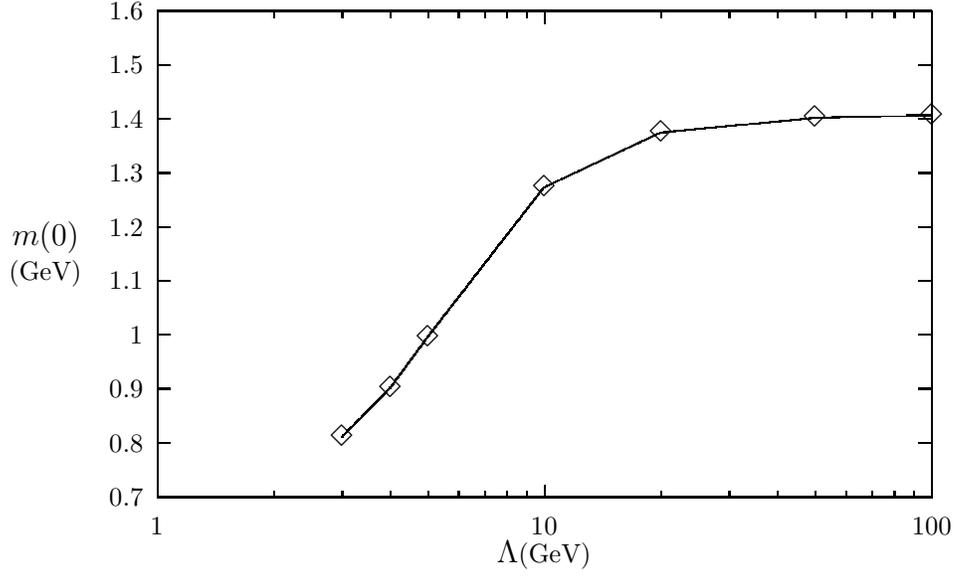
\begin{figure}[!htb]
\begin{center}
\setlength{\unitlength}{0.240900pt}
\ifx\plotpoint\undefined\newsavebox{\plotpoint}\fi
\sbox{\plotpoint}{\rule[-0.200pt]{0.400pt}{0.400pt}}%
\begin{picture}(1500,900)(0,0)
\font\gnuplot=cmr10 at 10pt
\gnuplot
\sbox{\plotpoint}{\rule[-0.200pt]{0.400pt}{0.400pt}}%
\put(220.0,113.0){\rule[-0.200pt]{4.818pt}{0.400pt}}
\put(198,113){\makebox(0,0)[r]{0.7}}
\put(1416.0,113.0){\rule[-0.200pt]{4.818pt}{0.400pt}}
\put(220.0,198.0){\rule[-0.200pt]{4.818pt}{0.400pt}}
\put(198,198){\makebox(0,0)[r]{0.8}}
\put(1416.0,198.0){\rule[-0.200pt]{4.818pt}{0.400pt}}
\put(220.0,283.0){\rule[-0.200pt]{4.818pt}{0.400pt}}
\put(198,283){\makebox(0,0)[r]{0.9}}
\put(1416.0,283.0){\rule[-0.200pt]{4.818pt}{0.400pt}}
\put(220.0,368.0){\rule[-0.200pt]{4.818pt}{0.400pt}}
\put(198,368){\makebox(0,0)[r]{1}}
\put(1416.0,368.0){\rule[-0.200pt]{4.818pt}{0.400pt}}
\put(220.0,453.0){\rule[-0.200pt]{4.818pt}{0.400pt}}
\put(198,453){\makebox(0,0)[r]{1.1}}
\put(1416.0,453.0){\rule[-0.200pt]{4.818pt}{0.400pt}}
\put(220.0,537.0){\rule[-0.200pt]{4.818pt}{0.400pt}}
\put(198,537){\makebox(0,0)[r]{1.2}}
\put(1416.0,537.0){\rule[-0.200pt]{4.818pt}{0.400pt}}
\put(220.0,622.0){\rule[-0.200pt]{4.818pt}{0.400pt}}
\put(198,622){\makebox(0,0)[r]{1.3}}
\put(1416.0,622.0){\rule[-0.200pt]{4.818pt}{0.400pt}}
\put(220.0,707.0){\rule[-0.200pt]{4.818pt}{0.400pt}}
\put(198,707){\makebox(0,0)[r]{1.4}}
\put(1416.0,707.0){\rule[-0.200pt]{4.818pt}{0.400pt}}
\put(220.0,792.0){\rule[-0.200pt]{4.818pt}{0.400pt}}
\put(198,792){\makebox(0,0)[r]{1.5}}
\put(1416.0,792.0){\rule[-0.200pt]{4.818pt}{0.400pt}}
\put(220.0,877.0){\rule[-0.200pt]{4.818pt}{0.400pt}}
\put(198,877){\makebox(0,0)[r]{1.6}}
\put(1416.0,877.0){\rule[-0.200pt]{4.818pt}{0.400pt}}
\put(220.0,113.0){\rule[-0.200pt]{0.400pt}{4.818pt}}
\put(220,68){\makebox(0,0){1}}
\put(220.0,857.0){\rule[-0.200pt]{0.400pt}{4.818pt}}
\put(403.0,113.0){\rule[-0.200pt]{0.400pt}{2.409pt}}
\put(403.0,867.0){\rule[-0.200pt]{0.400pt}{2.409pt}}
\put(510.0,113.0){\rule[-0.200pt]{0.400pt}{2.409pt}}
\put(510.0,867.0){\rule[-0.200pt]{0.400pt}{2.409pt}}
\put(586.0,113.0){\rule[-0.200pt]{0.400pt}{2.409pt}}
\put(586.0,867.0){\rule[-0.200pt]{0.400pt}{2.409pt}}
\put(645.0,113.0){\rule[-0.200pt]{0.400pt}{2.409pt}}
\put(645.0,867.0){\rule[-0.200pt]{0.400pt}{2.409pt}}
\put(693.0,113.0){\rule[-0.200pt]{0.400pt}{2.409pt}}
\put(693.0,867.0){\rule[-0.200pt]{0.400pt}{2.409pt}}
\put(734.0,113.0){\rule[-0.200pt]{0.400pt}{2.409pt}}
\put(734.0,867.0){\rule[-0.200pt]{0.400pt}{2.409pt}}
\put(769.0,113.0){\rule[-0.200pt]{0.400pt}{2.409pt}}
\put(769.0,867.0){\rule[-0.200pt]{0.400pt}{2.409pt}}
\put(800.0,113.0){\rule[-0.200pt]{0.400pt}{2.409pt}}
\put(800.0,867.0){\rule[-0.200pt]{0.400pt}{2.409pt}}
\put(828.0,113.0){\rule[-0.200pt]{0.400pt}{4.818pt}}
\put(828,68){\makebox(0,0){10}}
\put(828.0,857.0){\rule[-0.200pt]{0.400pt}{4.818pt}}
\put(1011.0,113.0){\rule[-0.200pt]{0.400pt}{2.409pt}}
\put(1011.0,867.0){\rule[-0.200pt]{0.400pt}{2.409pt}}
\put(1118.0,113.0){\rule[-0.200pt]{0.400pt}{2.409pt}}
\put(1118.0,867.0){\rule[-0.200pt]{0.400pt}{2.409pt}}
\put(1194.0,113.0){\rule[-0.200pt]{0.400pt}{2.409pt}}
\put(1194.0,867.0){\rule[-0.200pt]{0.400pt}{2.409pt}}
\put(1253.0,113.0){\rule[-0.200pt]{0.400pt}{2.409pt}}
\put(1253.0,867.0){\rule[-0.200pt]{0.400pt}{2.409pt}}
\put(1301.0,113.0){\rule[-0.200pt]{0.400pt}{2.409pt}}
\put(1301.0,867.0){\rule[-0.200pt]{0.400pt}{2.409pt}}
\put(1342.0,113.0){\rule[-0.200pt]{0.400pt}{2.409pt}}
\put(1342.0,867.0){\rule[-0.200pt]{0.400pt}{2.409pt}}
\put(1377.0,113.0){\rule[-0.200pt]{0.400pt}{2.409pt}}
\put(1377.0,867.0){\rule[-0.200pt]{0.400pt}{2.409pt}}
\put(1408.0,113.0){\rule[-0.200pt]{0.400pt}{2.409pt}}
\put(1408.0,867.0){\rule[-0.200pt]{0.400pt}{2.409pt}}
\put(1436.0,113.0){\rule[-0.200pt]{0.400pt}{4.818pt}}
\put(1436,68){\makebox(0,0){100}}
\put(1436.0,857.0){\rule[-0.200pt]{0.400pt}{4.818pt}}
\put(220.0,113.0){\rule[-0.200pt]{292.934pt}{0.400pt}}
\put(1436.0,113.0){\rule[-0.200pt]{0.400pt}{184.048pt}}
\put(220.0,877.0){\rule[-0.200pt]{292.934pt}{0.400pt}}
\put(45,495){\makebox(0,0){\shortstack{$m(0)$\\(GeV)}}}
\put(828,23){\makebox(0,0){$\Lambda$(GeV)}}
\put(220.0,113.0){\rule[-0.200pt]{0.400pt}{184.048pt}}
\put(510,207){\usebox{\plotpoint}}
\multiput(510.58,207.00)(0.499,0.506){149}{\rule{0.120pt}{0.505pt}}
\multiput(509.17,207.00)(76.000,75.951){2}{\rule{0.400pt}{0.253pt}}
\multiput(586.58,284.00)(0.499,0.678){115}{\rule{0.120pt}{0.642pt}}
\multiput(585.17,284.00)(59.000,78.667){2}{\rule{0.400pt}{0.321pt}}
\multiput(645.58,364.00)(0.500,0.645){363}{\rule{0.120pt}{0.616pt}}
\multiput(644.17,364.00)(183.000,234.722){2}{\rule{0.400pt}{0.308pt}}
\multiput(828.00,600.58)(1.066,0.499){169}{\rule{0.951pt}{0.120pt}}
\multiput(828.00,599.17)(181.026,86.000){2}{\rule{0.476pt}{0.400pt}}
\multiput(1011.00,686.58)(5.336,0.496){43}{\rule{4.309pt}{0.120pt}}
\multiput(1011.00,685.17)(233.057,23.000){2}{\rule{2.154pt}{0.400pt}}
\multiput(1253.00,709.60)(26.655,0.468){5}{\rule{18.400pt}{0.113pt}}
\multiput(1253.00,708.17)(144.810,4.000){2}{\rule{9.200pt}{0.400pt}}
\put(510,207){\raisebox{-.8pt}{\makebox(0,0){$\Diamond$}}}
\put(586,284){\raisebox{-.8pt}{\makebox(0,0){$\Diamond$}}}
\put(645,364){\raisebox{-.8pt}{\makebox(0,0){$\Diamond$}}}
\put(828,600){\raisebox{-.8pt}{\makebox(0,0){$\Diamond$}}}
\put(1011,686){\raisebox{-.8pt}{\makebox(0,0){$\Diamond$}}}
\put(1253,709){\raisebox{-.8pt}{\makebox(0,0){$\Diamond$}}}
\put(1436,713){\raisebox{-.8pt}{\makebox(0,0){$\Diamond$}}}
\end{picture}
\end{center}
\caption{Cut-off dependence of the effective gluon mass
($\alpha_s=0.4,\hspace{0.4cm} \sigma=0.18 GeV^2, \hspace{0.5cm} N_c=3$).}
\label{fig.3}
\end{figure}

Though the integral kernel of gap equation is obtained to the second order
in perturbation theory, the numerical solution of this equation
provides the gluon mass nonperturbatively. The linearized gap equation
is solved for $\omega({\bf k})$. The dispersion relation is displaced
in Fig. (\ref{fig.1}). The free behavior $\omega({\bf k})=k$
is recovered at high energy, and at low energy $\omega$ tends 
to the constituent gluon mass of roughly $0.9$ GeV.
This suggests the suitable parametrization
$\omega({\bf k})=k + m({\bf k})$, which defines 
the 'running' gluon mass $m({\bf k})$, Fig. (\ref{fig.2}).
A good fit to the numerical solution of the linearized gap equation is obtained by
\begin{eqnarray}
m({\bf k})=m(0){\rm exp}(-\frac{k}{\kappa})           
\,,\label{eq:4.2a}\end{eqnarray}
with an effective mass $m(0)=0.9$ GeV and $\kappa=0.95$ GeV.

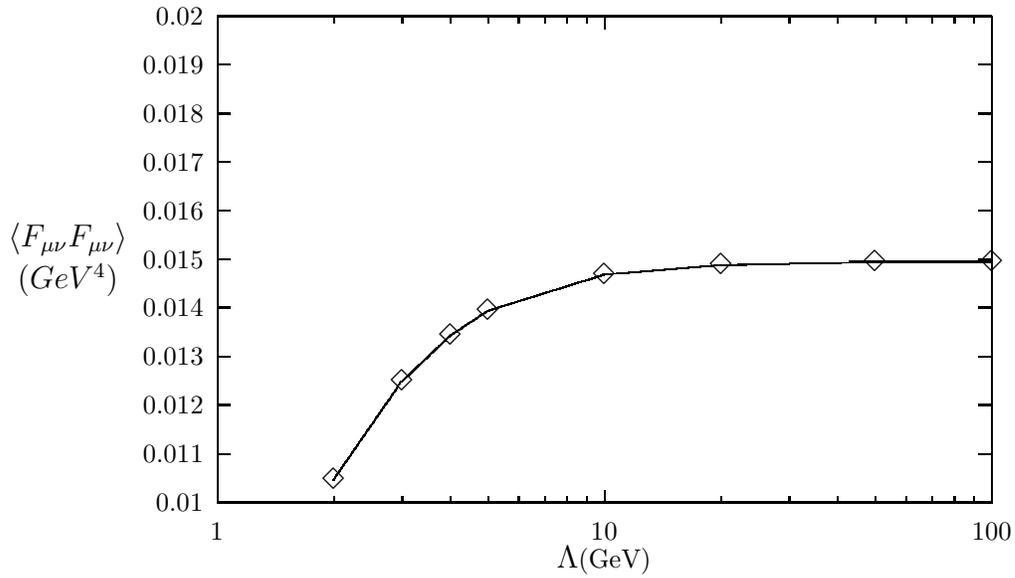
\begin{figure}[!htb]
\begin{center}
\setlength{\unitlength}{0.240900pt}
\ifx\plotpoint\undefined\newsavebox{\plotpoint}\fi
\begin{picture}(1500,900)(0,0)
\font\gnuplot=cmr10 at 10pt
\gnuplot
\sbox{\plotpoint}{\rule[-0.200pt]{0.400pt}{0.400pt}}%
\put(220.0,113.0){\rule[-0.200pt]{4.818pt}{0.400pt}}
\put(198,113){\makebox(0,0)[r]{0.01}}
\put(1416.0,113.0){\rule[-0.200pt]{4.818pt}{0.400pt}}
\put(220.0,189.0){\rule[-0.200pt]{4.818pt}{0.400pt}}
\put(198,189){\makebox(0,0)[r]{0.011}}
\put(1416.0,189.0){\rule[-0.200pt]{4.818pt}{0.400pt}}
\put(220.0,266.0){\rule[-0.200pt]{4.818pt}{0.400pt}}
\put(198,266){\makebox(0,0)[r]{0.012}}
\put(1416.0,266.0){\rule[-0.200pt]{4.818pt}{0.400pt}}
\put(220.0,342.0){\rule[-0.200pt]{4.818pt}{0.400pt}}
\put(198,342){\makebox(0,0)[r]{0.013}}
\put(1416.0,342.0){\rule[-0.200pt]{4.818pt}{0.400pt}}
\put(220.0,419.0){\rule[-0.200pt]{4.818pt}{0.400pt}}
\put(198,419){\makebox(0,0)[r]{0.014}}
\put(1416.0,419.0){\rule[-0.200pt]{4.818pt}{0.400pt}}
\put(220.0,495.0){\rule[-0.200pt]{4.818pt}{0.400pt}}
\put(198,495){\makebox(0,0)[r]{0.015}}
\put(1416.0,495.0){\rule[-0.200pt]{4.818pt}{0.400pt}}
\put(220.0,571.0){\rule[-0.200pt]{4.818pt}{0.400pt}}
\put(198,571){\makebox(0,0)[r]{0.016}}
\put(1416.0,571.0){\rule[-0.200pt]{4.818pt}{0.400pt}}
\put(220.0,648.0){\rule[-0.200pt]{4.818pt}{0.400pt}}
\put(198,648){\makebox(0,0)[r]{0.017}}
\put(1416.0,648.0){\rule[-0.200pt]{4.818pt}{0.400pt}}
\put(220.0,724.0){\rule[-0.200pt]{4.818pt}{0.400pt}}
\put(198,724){\makebox(0,0)[r]{0.018}}
\put(1416.0,724.0){\rule[-0.200pt]{4.818pt}{0.400pt}}
\put(220.0,801.0){\rule[-0.200pt]{4.818pt}{0.400pt}}
\put(198,801){\makebox(0,0)[r]{0.019}}
\put(1416.0,801.0){\rule[-0.200pt]{4.818pt}{0.400pt}}
\put(220.0,877.0){\rule[-0.200pt]{4.818pt}{0.400pt}}
\put(198,877){\makebox(0,0)[r]{0.02}}
\put(1416.0,877.0){\rule[-0.200pt]{4.818pt}{0.400pt}}
\put(220.0,113.0){\rule[-0.200pt]{0.400pt}{4.818pt}}
\put(220,68){\makebox(0,0){1}}
\put(220.0,857.0){\rule[-0.200pt]{0.400pt}{4.818pt}}
\put(403.0,113.0){\rule[-0.200pt]{0.400pt}{2.409pt}}
\put(403.0,867.0){\rule[-0.200pt]{0.400pt}{2.409pt}}
\put(510.0,113.0){\rule[-0.200pt]{0.400pt}{2.409pt}}
\put(510.0,867.0){\rule[-0.200pt]{0.400pt}{2.409pt}}
\put(586.0,113.0){\rule[-0.200pt]{0.400pt}{2.409pt}}
\put(586.0,867.0){\rule[-0.200pt]{0.400pt}{2.409pt}}
\put(645.0,113.0){\rule[-0.200pt]{0.400pt}{2.409pt}}
\put(645.0,867.0){\rule[-0.200pt]{0.400pt}{2.409pt}}
\put(693.0,113.0){\rule[-0.200pt]{0.400pt}{2.409pt}}
\put(693.0,867.0){\rule[-0.200pt]{0.400pt}{2.409pt}}
\put(734.0,113.0){\rule[-0.200pt]{0.400pt}{2.409pt}}
\put(734.0,867.0){\rule[-0.200pt]{0.400pt}{2.409pt}}
\put(769.0,113.0){\rule[-0.200pt]{0.400pt}{2.409pt}}
\put(769.0,867.0){\rule[-0.200pt]{0.400pt}{2.409pt}}
\put(800.0,113.0){\rule[-0.200pt]{0.400pt}{2.409pt}}
\put(800.0,867.0){\rule[-0.200pt]{0.400pt}{2.409pt}}
\put(828.0,113.0){\rule[-0.200pt]{0.400pt}{4.818pt}}
\put(828,68){\makebox(0,0){10}}
\put(828.0,857.0){\rule[-0.200pt]{0.400pt}{4.818pt}}
\put(1011.0,113.0){\rule[-0.200pt]{0.400pt}{2.409pt}}
\put(1011.0,867.0){\rule[-0.200pt]{0.400pt}{2.409pt}}
\put(1118.0,113.0){\rule[-0.200pt]{0.400pt}{2.409pt}}
\put(1118.0,867.0){\rule[-0.200pt]{0.400pt}{2.409pt}}
\put(1194.0,113.0){\rule[-0.200pt]{0.400pt}{2.409pt}}
\put(1194.0,867.0){\rule[-0.200pt]{0.400pt}{2.409pt}}
\put(1253.0,113.0){\rule[-0.200pt]{0.400pt}{2.409pt}}
\put(1253.0,867.0){\rule[-0.200pt]{0.400pt}{2.409pt}}
\put(1301.0,113.0){\rule[-0.200pt]{0.400pt}{2.409pt}}
\put(1301.0,867.0){\rule[-0.200pt]{0.400pt}{2.409pt}}
\put(1342.0,113.0){\rule[-0.200pt]{0.400pt}{2.409pt}}
\put(1342.0,867.0){\rule[-0.200pt]{0.400pt}{2.409pt}}
\put(1377.0,113.0){\rule[-0.200pt]{0.400pt}{2.409pt}}
\put(1377.0,867.0){\rule[-0.200pt]{0.400pt}{2.409pt}}
\put(1408.0,113.0){\rule[-0.200pt]{0.400pt}{2.409pt}}
\put(1408.0,867.0){\rule[-0.200pt]{0.400pt}{2.409pt}}
\put(1436.0,113.0){\rule[-0.200pt]{0.400pt}{4.818pt}}
\put(1436,68){\makebox(0,0){100}}
\put(1436.0,857.0){\rule[-0.200pt]{0.400pt}{4.818pt}}
\put(220.0,113.0){\rule[-0.200pt]{292.934pt}{0.400pt}}
\put(1436.0,113.0){\rule[-0.200pt]{0.400pt}{184.048pt}}
\put(220.0,877.0){\rule[-0.200pt]{292.934pt}{0.400pt}}
\put(45,495){\makebox(0,0){\shortstack{\hspace{-1cm}$\langle F_{\mu\nu}F_{\mu\nu} \rangle$\hspace{-0.4cm}\\ \hspace{-1cm}$(GeV^4)$}}}
\put(828,23){\makebox(0,0){$\Lambda$(GeV)}}
\put(220.0,113.0){\rule[-0.200pt]{0.400pt}{184.048pt}}
\put(403,148){\usebox{\plotpoint}}
\multiput(403.58,148.00)(0.499,0.725){211}{\rule{0.120pt}{0.679pt}}
\multiput(402.17,148.00)(107.000,153.590){2}{\rule{0.400pt}{0.340pt}}
\multiput(510.00,303.58)(0.527,0.499){141}{\rule{0.522pt}{0.120pt}}
\multiput(510.00,302.17)(74.916,72.000){2}{\rule{0.261pt}{0.400pt}}
\multiput(586.00,375.58)(0.758,0.498){75}{\rule{0.705pt}{0.120pt}}
\multiput(586.00,374.17)(57.536,39.000){2}{\rule{0.353pt}{0.400pt}}
\multiput(645.00,414.58)(1.611,0.499){111}{\rule{1.384pt}{0.120pt}}
\multiput(645.00,413.17)(180.127,57.000){2}{\rule{0.692pt}{0.400pt}}
\multiput(828.00,471.58)(6.243,0.494){27}{\rule{4.980pt}{0.119pt}}
\multiput(828.00,470.17)(172.664,15.000){2}{\rule{2.490pt}{0.400pt}}
\multiput(1011.00,486.59)(26.870,0.477){7}{\rule{19.460pt}{0.115pt}}
\multiput(1011.00,485.17)(201.610,5.000){2}{\rule{9.730pt}{0.400pt}}
\put(403,148){\raisebox{-.8pt}{\makebox(0,0){$\Diamond$}}}
\put(510,303){\raisebox{-.8pt}{\makebox(0,0){$\Diamond$}}}
\put(586,375){\raisebox{-.8pt}{\makebox(0,0){$\Diamond$}}}
\put(645,414){\raisebox{-.8pt}{\makebox(0,0){$\Diamond$}}}
\put(828,471){\raisebox{-.8pt}{\makebox(0,0){$\Diamond$}}}
\put(1011,486){\raisebox{-.8pt}{\makebox(0,0){$\Diamond$}}}
\put(1253,491){\raisebox{-.8pt}{\makebox(0,0){$\Diamond$}}}
\put(1436,491){\raisebox{-.8pt}{\makebox(0,0){$\Diamond$}}}
\put(1253.0,491.0){\rule[-0.200pt]{44.085pt}{0.400pt}}
\end{picture}
\end{center}
\caption{Gluon condensate
($\alpha_s=0.4, \sigma=0.18 GeV^2, N_c=3$).}
\label{fig.4}
\end{figure}

The gluon condensate is another nonperturbative characteristic calculated
within this approach. The condensate is given by the vacuum expectation value
$\langle 0|{\bf \Pi}^2 + {\bf B}^2|0 \rangle$ with only abelian component
for the magnetic field (pairing ansatz), Eq. (\ref{eq:2.9}). We regulate this
value by subtracting the perturbative contribution
\begin{eqnarray}   
\langle \frac{\alpha_s}{\pi}F^a_{\mu\nu}F^a_{\mu\nu} \rangle =
\frac{N_c^2-1}{\pi^3}\int_0^{\infty}dk k^2 \alpha_s
\frac{(\omega({\bf k})-k)^2}{2\omega({\bf k})}
\,,\label{eq:4.2a}\end{eqnarray}
Using the dispersion relation obtained above $\omega({\bf k})$
the gluon condensate is obtained $1.3\cdot 10^{-2} GeV^4 $ 
(for the cut-off $\Lambda=4$ GeV), that agrees with the sum rules \cite{ShVaZa}.
The dependence of the condensate on the cut-off is shown in Fig. (\ref{fig.4}).
The behavior of the effective gluon mass $m(0)$ with $\Lambda$ is displaced
in Fig. (\ref{fig.3}). Logarithmic dependence of both terms can be eliminated
by including the renormalization group running of coupling constant.

\subsection{Tamm-Dancoff approach for glueball spectrum}

We approximate a glueball bound state to consist of two constituent gluons.
The glueball wave function in the rest frame is 
\begin{eqnarray}
|\psi_n\rangle=\int{d{\bf q}\over(2\pi)^3}\phi_n^{ij}({\bf q})
a_i^{a\dagger}({\bf q})a_j^{a\dagger}(-{\bf q})|0\rangle
\,,\label{eq}\end{eqnarray}
where $a^{\dagger}$ is a creation operator of quasiparticle
with an effective dispersion relation $\omega_{\bf q}$.
Tamm-Dancoff approximation works reasonable for a block-diagonal
effective Hamiltonian. To obtain the Tamm-Dancoff bound state equation
we project the Schr{\"o}dinger equation 
$H|\psi_n\rangle=E_n|\psi_n\rangle$
on the two-body sector; the result reads
\begin{eqnarray} 
\langle\psi_n|[H,a^{i\dagger}_{\bf q}a^{j\dagger}_{-{\bf q}}]|0\rangle
=(E_n-E_0)\left(X^{ij}_{i'j'}({\bf q})\phi_n^{i'j'}({\bf q}) \right)
\,,\label{eq:4.6def}\end{eqnarray}
with the notation
\begin{eqnarray}
\left(X^{ij}_{i'j'}({\bf q})\phi_n^{i'j'}({\bf q}) \right) 
= D_{ii'}({\bf q})D_{jj'}({\bf q})\phi_n^{i'j'}({\bf q})
+ D_{ij'}({\bf q})D_{ji'}({\bf q})\phi_n^{i'j'}(-{\bf q})
\,,\label{eq:4.6}\end{eqnarray}
and $a^{i\dagger}_{\bf q}=a^{i\dagger}({\bf q})$, color indices are omitted.
In Eq. (\ref{eq:4.6}) we subtract the trivial vacuum energy $E_0$,
defined as $H|0\rangle=E_0|0\rangle$. Tamm-Dancoff equation with all possible
terms obtained to the second order is given in Appendix (\ref{A}).
We consider here the most essential part of the integral kernel,
neglecting the perturbative transverse gluon exchange and terms
coming from the normal-ordering of four-gluon vertex.
The contribution of the neglected terms to the mass spectrum
is expected to be small.
Tamm-Dancoff equation for (pseudo)scalar glueball states 
(with total angular momentum $J$, parity $P$, and charge conjugation $C$) 
reads
\begin{eqnarray} 
& & (E_n-E_0)\phi_n(q)
= \left[\left(\frac{q^2+\widetilde{m}_{CT}^2}
{\omega_{\bf q}}+\omega_{\bf q} \right)\right.
\nonumber\\
&+& \left.\frac{\alpha_s}{4} N_c
\int{d{\bf p}\over(2\pi)^3}
V({\bf p}-{\bf q})\left(1+({\hat p}{\hat q})^2 \right)
\frac{\omega_{\bf p}^2+\omega_{\bf q}^2}{\omega_{\bf p}\omega_{\bf q}}
{\rm e}^{-p^2/\Lambda^2}
\right]\phi_n(q)
\nonumber\\
&-& \frac{\alpha_s}{8} N_c \int{d{\bf p}\over(2\pi)^3}V({\bf p}-{\bf q})
\frac{(\omega_{\bf p}+\omega_{\bf q})^2}{\omega_{\bf p}\omega_{\bf q}}
2({\hat p}{\hat q})F^{JPC}({\bf p},{\bf q})\phi_n(p)
\,,\label{eq:4.7}\end{eqnarray}
with
\begin{eqnarray}
F^{0++}({\bf p},{\bf q}) &=& 1+({\hat p}{\hat q})^2
\nonumber\\
F^{0-+}({\bf p},{\bf q}) &=& 2({\hat p}{\hat q})
\,,\label{eq:4.8}\end{eqnarray}
and the Coulomb counterterm $\widetilde{m}$ given by Eq. (\ref{eq:4.2ct}).

Bound state equation (\ref{eq:4.7})
has two types of divergences, UV associated with canonical Coulomb interaction
and IR coming from confinement. UV divergences in the self-energy term
are regulated by the cut-off function, and the leading divergent part
is canceled by the mass counterterm $\widetilde{m}$. UV behavior
of potential part (the last term) is regulated by the wave function.
Though kinetic (self energy term) and potential parts contribute the infrared 
divergent pieces the bound state equation is infrared finite 
due to complete cancelation of both parts in infrared region.
This cancelation  happens for the color-singlet objects \cite{adler}.

Numerical calculations of Tamm-Dancoff equation (\ref{eq:4.7})
are performed variationally with a set of gaussian test functions.
Results of calculations for the lowest glueball states are presented
in Table (\ref{tab.1}) and compared with the available lattice data \cite{lattice}.

\begin{table}[ht]
$$
\begin{tabular}{|c|c|c|c|c|} \hline 
  $J^{PC}$ & $0^{++}$ & $0^{*++}$ & $0^{-+}$ & $0^{*-+}$ \\ \hline
 Tamm-Dancoff, (MeV) & $1760$ & $2697$ 
& $2142$ & $2895$ \\ \hline
lattice data \cite{lattice}, (MeV) & $1730(80)$ & $2670(130)$   
& $2590(130)$ & $3640(180)$ \\ \hline 
\end{tabular}
$$
\caption{Glueball spectrum for the lowest scalar and pseudoscalar states
($\alpha_s=0.4, \sigma=0.18 GeV^2, \Lambda=4 GeV, N_c=3$).}
\label{tab.1}
\end{table}
 
Lattice calculations are done for $SU(3)$ pure gluodynamics, 
using anisotropic lattice and improved SII action.
Better agreement with the lattice data is achieved for the scalar channel.
Remarkable, the mass of the lowest scalar glueball $0^{++}$
is roughly twice of the effective gluon mass $m(0)$ obtained before.

\section{Conclusion}

We outlined the systematic procedure to solve the bound state problem in QCD.
Our study is driven by the success of the constituent quark model.
The approach is based on the method of flow equations,
that includes the renormalization group scaling of canonical Hamiltonian
from high to low energies and simultaneous change of basis
from current to constituent degrees of freedom.
The renormalization group aspect of the method protects the observables,
calculated from effective Hamiltonian to be practically free from UV divergences.
Therefore we are able to include radiative corrections 
at one-loop level. The leading UV behavior is canceled by the mass counterterm.
The slow logarithmic dependence of masses on the cut-off, which is still left
after perturbative renormalization, can be absorbed by the nonperturbative renormalization  
of strong coupling constant. Confinement interaction can cause the infrared problem.
In glueball bound state equation this potential problem is avoided
by adding a diagram with soft gluon emission and absorbtion in the second order.
This leads to complete cancelation of infrared divergent parts for color singlet objects.

The use of effective basis (built by including all possible particle number conserving 
interactions in $H_d$ and acting in QCD vacuum) in conjunction with 
the renormalization group scaling makes possible to get
an effective block-diagonal Hamiltonian with fixed number of quasiparticles
(constituent gluons).
An effective Hamiltonian provides a constituent description for glueball bound states
at low energies. 

Based on flow equations, we are able to extend the study to nonperturbative range of energies.
The renormalized effective Hamiltonian, obtained in perturbative frame by flow equations
and including confining interaction, can be represented in the nonperturbative 
(effective) Fock space provided the canonical form. Contrary, the effective Hamiltonian,
obtained in similarity scheme, is band-diagonal only in perturbative frame
and can not be transformed in nonperturbative basis.
Though coupling constant is not small at low energies, $\alpha_s\sim 1$,
we are still able to perform analytical calculations by flow equations,
since higher orders are suppressed by a (heavy) effective gluon mass.
    
To provide a detailed analyses and to compare with lattice data we solve numerically
the gap equation for an effective gluon energy and Tamm-Dancoff glueball bound state equation.
Roughly the glueball mass is twice of the effcetive gluon mass,
that supports the constituent picture. The gluon condensate based
on the effective gluon dispersion relation is in agreement
with QCD sum rules.

\vspace{1cm}

{\bf Acknowledgments.}\thinspace One of the authors (E.G.) is thankful 
to Dave Robertson for materials provided and help, and also 
to Felipe Llanes-Estrada for useful discussion and reading the manuscipt. 
Also E.G. would like to thank Karen Avetovich Ter-Martirosyan for discussions 
and introducing the idea. 
This work was supported by the U.S. DOE under contracts
DE-FGO2-96ER40947. The North Carolina Supercomputing Center and
the National Energy Research Scientific Computer Center are also acknowledged
for the grant of supercomputer time.

\newpage

\newpage

\newpage
\appendix
\section{The complete form of Tamm-Dancoff equation}
\label{A}

We outline the complete form of Tamm-Dancoff equation
with all possible terms obtained to the second order.
One- (only $a^{\dagger}a$ component) and two-body sectors,
specified in  Eq. (\ref{eq:3.37d}),    
contribute to the color singlet glueball state. 
Making use of Eq. (\ref{eq:4.6def}), we get
the resulting Tamm-Dancoff equation for glueball bound state 
\begin{eqnarray} 
& & (E_n-E_0)\left( X^{ij}_{ln}({\bf q})\phi_{+}^{ln}({\bf q})
+Y^{ij}_{ln}({\bf q})\phi_{-}^{ln}({\bf q}) \right)
= \left[\left(\frac{q^2+m_{CT}^2}{\omega_{\bf q}}+\omega_{\bf q} \right)\right.
\nonumber\\
&+& \left.\frac{\alpha_s}{4} N_c
\int{d{\bf p}\over(2\pi)^3}
V({\bf p}-{\bf q})\left(1+({\hat p}{\hat q})^2 \right)
\frac{\omega_{\bf p}^2+\omega_{\bf q}^2}{\omega_{\bf p}\omega_{\bf q}}
f^2(\omega_{\bf p};\Lambda) \right.
\nonumber\\
&+&\left. \alpha_s\pi N_c
\int{d{\bf p}\over(2\pi)^3}{1\over\omega_{\bf p}\omega_{\bf q}}
\left(3-({\hat k}{\hat q})^2 \right)
f^2(\omega_{\bf p};\Lambda) \right.
\nonumber\\
&+&\left. \alpha_s\pi N_c
\int{d{\bf p}\over(2\pi)^3}{1\over\omega_{\bf p}\omega_{\bf q}\omega_{{\bf p}-{\bf q}} }
G({\bf p},{\bf q})\left(\frac{f^2(D_0;\Lambda)}{D_0}+\frac{f^2(D_1;\Lambda)}{D_1} \right)
\right]
\nonumber\\
&\times&\left( X^{ij}_{ln}({\bf q})\phi_{+}^{ln}({\bf q})
+Y^{ij}_{ln}({\bf q})\phi_{-}^{ln}({\bf q})
\right)
\nonumber\\
&+&\left[
-\frac{\alpha_s}{8} N_c \int{d{\bf p}\over(2\pi)^3}V({\bf p}-{\bf q})
\frac{(\omega_{\bf p}+\omega_{\bf q})^2}{\omega_{\bf p}\omega_{\bf q}}
\left( X^{ij}_{km}({\bf q})X^{km}_{ln}({\bf p})\phi_{+}^{ln}({\bf p})
+Y^{ij}_{km}({\bf q})Y^{km}_{ln}({\bf p})\phi_{-}^{ln}({\bf p})
\right) \right.
\nonumber\\
&+&\left.
\alpha_s\pi N_c \int{d{\bf p}\over(2\pi)^3}
\frac{2}{\omega_{\bf p}\omega_{\bf q}}
\left( D_{ij}({\bf q})D_{ln}({\bf p})
-\frac{1}{4}X^{ij}_{km}({\bf q})X^{km}_{ln}({\bf p}) \right) \phi_{+}^{ln}({\bf p})
\right.
\nonumber\\
&-&\left.
\alpha_s 2\pi N_c\int{d{\bf p}\over(2\pi)^3}
{1\over\omega_{\bf p}\omega_{\bf q}\omega_{{\bf p}-{\bf q}} }
\left(\frac{\Theta(D_1,D_{1'})}{D_1}+\frac{\Theta(D_{1'},D_{1})}{D_{1'}} \right)
\right.
\nonumber\\
&\times&\left.S_{km,k'm'}({\bf q},{\bf p})
\left( X^{ij}_{kk'}({\bf q})X^{mm'}_{ln}({\bf p})\phi_{+}^{ln}({\bf p})
+Y^{ij}_{kk'}({\bf q})Y^{mm'}_{ln}({\bf p})\phi_{-}^{ln}({\bf p}) \right)
\right]
\,,\label{eq}\end{eqnarray}
where the energy denominators are
\begin{eqnarray}
D_0 &=&-(\omega_{\bf q}+\omega_{\bf p}+\omega_{{\bf p}-{\bf q}}) \nonumber\\
D_1 &=&-(-\omega_{\bf q}+\omega_{\bf p}+\omega_{{\bf p}-{\bf q}}) \nonumber\\
D_{1'} &=&-(-\omega_{\bf p}+\omega_{\bf q}+\omega_{{\bf p}-{\bf q}}) 
\,,\label{eq}\end{eqnarray}
and the following notations were introduced
\begin{eqnarray}    
&& \phi_{+}^{ln}({\bf p})=\frac{1}{2}(\phi^{ln}({\bf p})+\phi^{ln}(-{\bf p}) )\nonumber\\
&& \phi_{-}^{ln}({\bf p})=\frac{1}{2}(\phi^{ln}({\bf p})-\phi^{ln}(-{\bf p}) )\nonumber\\
&& X^{ij}_{ln}({\bf q})=D_{il}({\bf q})D_{jn}({\bf q})+D_{in}({\bf q})D_{jl}({\bf q})\nonumber\\
&& Y^{ij}_{ln}({\bf q})=D_{il}({\bf q})D_{jn}({\bf q})-D_{in}({\bf q})D_{jl}({\bf q})
\,,\label{eq}\end{eqnarray}
Canonical UV divergencies are regulated by the cut-off function,
that is specified in the main text; factor $S_{km,k'm'}$ is given in Eq. (\ref{eq:3.15}),
and $\Theta$ factors in Eq. (\ref{eq:3.19a}). 
We write Tamm-Dancoff equation explicitly in the scalar ($0^{++}$)
and pseudoscalar ($0^{-+}$) channels.
For $0^{++}$ the wave function can be written
$\phi^{ij}=\delta^{ij}\phi(q)$. Tamm-Dancoff equation takes the form
\begin{eqnarray} 
& & (E_n-E_0)\phi_n(q)=
 \left[\left(\frac{q^2+m_{CT}^2}{\omega_{\bf q}}+\omega_{\bf q} \right)\right.
\nonumber\\
&+& \left.\frac{\alpha_s}{4} N_c
\int{d{\bf p}\over(2\pi)^3}
V({\bf p}-{\bf q})\left(1+({\hat p}{\hat q})^2 \right)
\frac{\omega_{\bf p}^2+\omega_{\bf q}^2}{\omega_{\bf p}\omega_{\bf q}}
f^2(\omega_{\bf p};\Lambda)\right.
\nonumber\\
&+&\left. \alpha_s\pi N_c
\int{d{\bf p}\over(2\pi)^3}{1\over\omega_{\bf p}\omega_{\bf q}}
\left(3-({\hat k}{\hat q})^2 \right)
f^2(\omega_{\bf p};\Lambda)\right.
\nonumber\\
&+&\left. \alpha_s\pi N_c
\int{d{\bf p}\over(2\pi)^3}{1\over\omega_{\bf p}\omega_{\bf q}\omega_{{\bf p}-{\bf q}} }
G({\bf p},{\bf q})\left(\frac{f^2(D_0;\Lambda)}{D_0}+\frac{f^2(D_1;\Lambda)}{D_1} \right)
\right]\thinspace \phi_n(q)
\nonumber\\
&+&\left[
-\frac{\alpha_s}{8} N_c \int{d{\bf p}\over(2\pi)^3}V({\bf p}-{\bf q})
\frac{(\omega_{\bf p}+\omega_{\bf q})^2}{\omega_{\bf p}\omega_{\bf q}}
(1+({\hat p}{\hat q})^2)\phi_n(p) \right.
\nonumber\\
&+&\left.
\alpha_s\pi N_c \int{d{\bf p}\over(2\pi)^3}
\frac{1}{2\omega_{\bf p}\omega_{\bf q}}
(3-({\hat p}{\hat q})^2)\phi_n(p) \right.
\nonumber\\
&-&\left.
\alpha_s 2\pi N_c\int{d{\bf p}\over(2\pi)^3}
{1\over\omega_{\bf p}\omega_{\bf q}\omega_{{\bf p}-{\bf q}} }
\left(\frac{\Theta(D_1,D_{1'})}{D_1}+\frac{\Theta(D_{1'},D_{1})}{D_{1'}} \right)
G({\bf q},{\bf p})\phi_n(p) \right]
\,,\label{eq:a.1}\end{eqnarray}
In the pseudoscalar channel, $0^{-+}$ glueball state, the wave function can be represented\\
$\phi^{ij}({\bf q})=\epsilon^{ijk}{\hat q}^{k}\phi(q)$.
Tamm-Dancoff equation reads
\begin{eqnarray} 
& & (E_n-E_0)\phi_n(q)
= \left[\left(\frac{q^2+m_{CT}^2}{\omega_{\bf q}}+\omega_{\bf q} \right)\right.
\nonumber\\
&+& \left.\frac{\alpha_s}{4} N_c
\int{d{\bf p}\over(2\pi)^3}
V({\bf p}-{\bf q})\left(1+({\hat p}{\hat q})^2 \right)
\frac{\omega_{\bf p}^2+\omega_{\bf q}^2}{\omega_{\bf p}\omega_{\bf q}}
f^2(\omega_{\bf p};\Lambda)\right.
\nonumber\\
&+&\left. \alpha_s\pi N_c
\int{d{\bf p}\over(2\pi)^3}{1\over\omega_{\bf p}\omega_{\bf q}}
\left(3-({\hat k}{\hat q})^2 \right)
f^2(\omega_{\bf p};\Lambda)\right.
\nonumber\\
&+&\left. \alpha_s\pi N_c
\int{d{\bf p}\over(2\pi)^3}{1\over\omega_{\bf p}\omega_{\bf q}\omega_{{\bf p}-{\bf q}} }
G({\bf q},{\bf p})\left(\frac{f^2(D_0;\Lambda)}{D_0}+\frac{f^2(D_1;\Lambda)}{D_1} \right)
\right]\phi_n(q)
\nonumber\\
&+&\left[
-\frac{\alpha_s}{8} N_c \int{d{\bf p}\over(2\pi)^3}V({\bf p}-{\bf q})
\frac{(\omega_{\bf p}+\omega_{\bf q})^2}{\omega_{\bf p}\omega_{\bf q}}
2({\hat p}{\hat q})\phi_n(p)\right.
\nonumber\\
&-&\left.
\alpha_s 2\pi N_c\int{d{\bf p}\over(2\pi)^3}
{1\over\omega_{\bf p}\omega_{\bf q}\omega_{{\bf p}-{\bf q}} }
\left(\frac{\Theta(D_1,D_{1'})}{D_1}+\frac{\Theta(D_{1'},D_{1})}{D_{1'}} \right)
\right.
\nonumber\\
&\times&\left.
2p^2q^2(1-({\bf p}{\bf q})^2)\left(\frac{2}{pq}
+\frac{1}{({\bf p}-{\bf q})^2} ({\bf p}{\bf q})\right)\phi_n(p)
\right]
\,,\label{eq:a.2}\end{eqnarray}
The reduced version of Eq. (\ref{eq:a.1}) and Eq. (\ref{eq:a.2})
with the Coulomb term in the integral kernel is used in the main text 
for numerical calculations.

\newpage

%

\end{document}